\documentclass{emulateapj}
\usepackage{graphicx}
\usepackage[space]{grffile}
\usepackage{latexsym}
\usepackage{textcomp}
\usepackage{longtable}
\usepackage{multirow,booktabs}
\usepackage{amsfonts,amsmath,amssymb}
\usepackage{natbib}
\usepackage{url}
\usepackage{CJKutf8}
\usepackage{hyperref}
\usepackage{enumerate}
\hypersetup{colorlinks=false,pdfborder={0 0 0}}
\newif\iflatexml\latexmlfalse
\usepackage[utf8]{inputenc}
\usepackage[english]{babel}
\usepackage{graphicx, color}

\newcommand\T{\rule{0pt}{2.6ex}}       

\newcommand{\be}{\begin{equation}}
\newcommand{\ee}{\end{equation}}

\definecolor{todo}{RGB}{200,0,0}

\shorttitle{Data Release 15}
\shortauthors{SDSS Collaboration}

\begin{document}

\title{The Fifteenth Data Release of the Sloan Digital Sky Surveys:  First Release of MaNGA Derived Quantities, Data Visualization Tools and Stellar Library}

\email{spokesperson@sdss.org}
\author{D. S. Aguado\altaffilmark{1},
Romina Ahumada\altaffilmark{2},
Andr\'es Almeida\altaffilmark{3},
Scott F. Anderson\altaffilmark{4},
Brett H. Andrews\altaffilmark{5},
Borja Anguiano\altaffilmark{6},
Erik Aquino Ort\'iz\altaffilmark{7},
Alfonso Arag\'on-Salamanca\altaffilmark{8},
Maria Argudo-Fern{\'a}ndez\altaffilmark{9,10},
Marie Aubert\altaffilmark{11}
% ...
}
\altaffiltext{1}{Instituto de Astrof\'isica de Canarias, E-38205 La Laguna, Tenerife, Spain}
\altaffiltext{2}{Departamento de Astronomia, Casilla 160-C, Universidad de Concepcion, Concepcion, Chile}
\altaffiltext{3}{Instituto de Investigaci\'{o}n Multidisciplinario en Ciencia y Tecnolog\'ia, Universidad de La Serena, Benavente 980, La Serena, Chile}
\altaffiltext{4}{Department of Astronomy, Box 351580, University of Washington, Seattle, WA 98195, USA}
\altaffiltext{5}{PITT PACC, Department of Physics and Astronomy, University of Pittsburgh, Pittsburgh, PA 15260, USA}
\altaffiltext{6}{Department of Astronomy, University of Virginia, 530 McCormick Road, Charlottesville, VA 22904-4325, USA}
\altaffiltext{7}{Instituto de Astronom{\'i}a, Universidad Nacional Aut\'onoma de M\'exico, A.P. 70-264, 04510, M\'exico, D.F., M\'exico}
\altaffiltext{8}{School of Physics \& Astronomy, University of Nottingham, Nottingham, NG7 2RD, UK}
\altaffiltext{9}{Centro de Astronom\'ia (CITEVA), Universidad de Antofagasta, Avenida Angamos 601 Antofagasta, Chile}
\altaffiltext{10}{Chinese Academy of Sciences South America Center for Astronomy, China-Chile Joint Center for Astronomy, Camino El Observatorio 1515, Las Condes, Santiago, Chile}
\altaffiltext{11}{Aix Marseille Univ, CNRS/IN2P3, CPPM, Marseille, France}
% ...

\begin{abstract}
Twenty years have passed since first light for the Sloan Digital Sky Survey (SDSS). Here, we release data taken by the fourth phase of SDSS (SDSS-IV) across its first three years of operation (July 2014--July 2017). This is the third data release for SDSS-IV, and the fifteenth from SDSS (Data Release Fifteen; DR15). New data come from MaNGA -- we release $4824$ datacubes, as well as the first stellar spectra in the MaNGA Stellar Library (MaStar), the first set of survey-supported analysis products (e.g. stellar and gas kinematics, emission line, and other maps) from the MaNGA Data Analysis Pipeline (DAP), and a new data visualisation and access tool we call ``Marvin". The next data release, DR16, will include new data from both APOGEE-2 and eBOSS; those surveys release no new data here, but we document updates and corrections to their data processing pipelines. The release is cumulative; it also includes the most recent reductions and calibrations of all data taken by SDSS since first light. In this paper we describe the location and format of the data and tools and cite technical references describing how it was obtained and processed. The SDSS website (www.sdss.org) has also been updated, providing links to data downloads, tutorials and examples of data use. While SDSS-IV will continue to collect astronomical data until 2020, and will be followed by SDSS-V (2020--2025), we end this paper by describing plans to ensure the sustainability of the SDSS data archive for many years beyond the collection of data.
\end{abstract}


\section{Introduction}
\label{sec:intro}
\setcounter{footnote}{0}

The Sloan Digital Sky Survey (SDSS; \citealt{2000AJ....120.1579Y}) data releases began with the Early Data
Release, or EDR, in June 2001 (\citealt{2002AJ....123..485S}) and have
been heavily used by astronomers and the broader public since that
time \citep{Raddick2014a,Raddick2014b}. Here we present the fifteenth public data release from SDSS, or DR15, made publicly available on 10th December 2018. 

SDSS has been marked by four phases so far,
with plans for a fifth.  Details are available in the papers
describing SDSS-I (EDR, DR1--DR5; \citealt{2000AJ....120.1579Y}),
SDSS-II (DR6--DR7; \citealt{frieman08b, 2009AJ....137.4377Y}),
SDSS-III (DR8--DR12; \citealt{2011AJ....142...72E}), and SDSS-IV
(DR13--DR15;
\citealt{2017AJ....154...28B}). \citet{2017arXiv171103234K} describe
the plans for SDSS-V, to start in mid-2020.

The data releases contain information about SDSS optical broad band
imaging, optical spectroscopy, and infrared spectroscopy. Currently,
SDSS-IV conducts optical and infrared spectroscopy (using two dedicated spectrographs; \citealt{Smee2013,Wilson2018}) at the 2.5-meter Sloan Foundation Telescope at Apache Point Observatory (APO; \citealt{2006AJ....131.2332G})
and infrared spectroscopy at the du Pont Telescope at Las Campanas
Observatory (LCO; \citealt{bowen73a}).

SDSS-IV began observations in July 2014, and
consists of three programs:
\begin{enumerate}
\item The extended Baryon Oscillation Spectroscopic Survey (eBOSS;
  \citealt{2016AJ....151...44D}) is surveying galaxies and quasars at
  redshifts $z\sim 0.6$--$3.5$ for large scale structure. It includes
  two major subprograms:
\begin{itemize}
  \item SPectroscopic IDentification of ERosita Sources (SPIDERS; \citealt{2017MNRAS.469.1065D})
  investigates the nature of $X$-ray emitting sources, including
  active galactic nuclei and galaxy clusters. 
\item Time Domain Spectroscopic Survey (TDSS; \citealt{Morganson2015})
  is exploring the physical nature of time-variable sources through
  spectroscopy.
\end{itemize}
\item Mapping Nearby Galaxies at APO (MaNGA;
  \citealt{2015ApJ...798....7B}) uses integral field spectroscopy
  (IFS) to study a representative sample of $\sim$10,000 nearby galaxies.
\item APOGEE-2 (the second phase of the APO Galactic Evolution Experiment or APOGEE;  \citealt{Majewski2017}) performs a large-scale and systematic investigation of the entire Milky Way Galaxy with near-infrared, high-resolution, and
  multiplexed instrumentation.
\end{enumerate}

SDSS-IV has had two previous data releases (DR13 and DR14;
\citealt{2017ApJS..233...25A, 2018ApJS..235...42A} respectively), containing the
  first two years of eBOSS, MaNGA, and APOGEE-2 data and new
  calibrations of the SDSS imaging data set.

DR15 contains new reductions and new data from MaNGA. This release includes the
first three years of MaNGA data plus a new suite of derived data
products based on the MaNGA data cubes, a new data access
tool for MaNGA known as Marvin, and data from a large ancillary programme aimed at improving the stellar library available for MaNGA (MaStar; the MaNGA Stellar Library). 

The full scope of the data release is described in Section 2, and
information on data distribution is given in Section 3. Each of the
sub-surveys is described in its own section, with MaNGA in Section 4 and APOGEE-2 and eBOSS (including
SPIDERS and TDSS) in Section 5.1 and 5.2, respectively. We discuss future plans for SDSS-IV and beyond in Section
6. Readers wanting a glossary of terms and acronyms used in SDSS can find one at \url{https://www.sdss.org/dr15/help/glossary/}.

\section{Scope of Data Release 15}
\label{sec:scope}

\begin{deluxetable*}{lrrr}
\tablewidth{5in}
\tablecaption{Reduced SDSS-IV spectroscopic data in DR15 \label{table:scope}} 
\tablehead{ 
\colhead{Target Category} & \colhead{\# DR13} & \colhead{\# DR13+14}  & \colhead{\# DR13+14+15}}
\startdata
\multicolumn{4}{l}{eBOSS} \\ 
\multicolumn{1}{r}{LRG samples} & 32968 & 138777 & 138777 \\
\multicolumn{1}{r}{ELG Pilot Survey} & 14459 & 35094 & 35094  \\
\multicolumn{1}{r}{Main QSO Sample}  & 33928 & 188277 & 188277 \\	
\multicolumn{1}{r}{Variability Selected QSOs} & 22756 & 87270 & 87270  \\
\multicolumn{1}{r}{Other QSO samples} & 24840 & 43502  & 43502  \\
\multicolumn{1}{r}{TDSS Targets} & 17927 & 57675 & 57675  \\
\multicolumn{1}{r}{SPIDERS Targets} &  3133 & 16394 & 16394  \\
\multicolumn{1}{r}{Standard Stars/White Dwarfs} &  53584 & 63880 & 63880  \\
\tableline 
\multicolumn{4}{l}{APOGEE-2} \T \\
\multicolumn{1}{r}{All Stars}  & 164562 & 263444 & 263444 \\
\multicolumn{1}{r}{NMSU 1-meter stars}  & 894 & 1018 & 1018 \\
\multicolumn{1}{r}{Telluric stars} & 17293 &  27127 &  27127 \\
\multicolumn{1}{r}{APOGEE-N Commissioning stars}  & 11917  & 12194 & 12194 \\
\tableline 
\multicolumn{1}{l}{MaNGA Cubes} &  1390 &  2812 &  4824 \\ 
\multicolumn{4}{l}{MaNGA main galaxy sample: } \\ 
\multicolumn{1}{r}{\tt PRIMARY\_v1\_2} &  600  & 1278  & 2126 \\ 
\multicolumn{1}{r}{\tt SECONDARY\_v1\_2} &  473  & 947  & 1665 \\ 
\multicolumn{1}{r}{\tt COLOR-ENHANCED\_v1\_2} & 216  & 447 & 710 \\  
\multicolumn{1}{l}{MaStar (MaNGA Stellar Library)} & 0 & 0 & 3326 \\
\multicolumn{1}{l}{Other MaNGA ancillary targets\tablenotemark{1}} &  31 & 121 & 324\\
\vspace{-0.1in}
\tablenotetext{1}{Many MaNGA ancillary targets were also observed as
  part of the main galaxy sample, and are counted twice in this table; some ancillary targets are not galaxies.} 
\enddata
\end{deluxetable*}

As with all previous SDSS public data releases, DR15 is cumulative and includes all data products that have been publicly released in earlier SDSS data releases. All previous releases are archived online to facilitate science replication; however we recommend new users always make use of the latest DR (even when using older data) to ensure they are using the most recent reduction routines. The scope of DR15 is shown in Table~\ref{table:scope}, and its components can be summarized as follows.

\begin{enumerate}
\item MaNGA integral-field spectroscopic data from 285 plates, including 119 plates observed between 26 September 2016 (MJD 57658) and 29 June 2017 (MJD 57934) that are newly released data in DR15. This data set is identical to the internally released MaNGA Product Launch-7 (MPL-7), and contains the same set of galaxies but processed with a different version of the reduction pipeline as the earlier internally released MPL-6. DR15 contains 4824 reconstructed 3D data cubes, of which 4688 are target galaxies (the remainder are ancillary targets which include galaxies, parts of galaxies, and some deep sky fields). This dataset includes 67 repeat observations, so that the total number of unique galaxies in this data release is 4621. Most of these galaxies are part of the MaNGA main sample, but ancillary target galaxies are also included in this count (see Table \ref{table:ancillary} for a summary of these). 
\item In addition to the MaNGA data cubes, DR15 also releases for the first time data products generated by the Data Analysis Pipeline (see section~\ref{sec:mangadap}). These products are available for all data cubes in DR15, with the exception of the cubes generated by some ancillary programs (i.e., Coma, IC342 and M31) if they do not have redshifts (e.g. sky fields). 
\item Alongside the new MaNGA data, and data products, DR15 also marks the launch of Marvin: a new tool to visualize and analyze MaNGA datacubes and maps (see section~\ref{sec:marvin}).
\item DR15 is the first public data release for the MaNGA Stellar Library MaStar (see section~\ref{sec:mastar}), containing 3326 optical stellar spectra.
\item In addition to updates to two previously released Value Added Catalogs (VACs), DR15 also includes six new VACs contributed by the MaNGA team (see Table~\ref{table:vac}). This brings the total number of VACs in the SDSS public data releases to 40.
\item Finally, DR15 includes a re-release of all previous versions of SDSS data releases. This includes the most recent data releases for APOGEE-2 and eBOSS \citep[described in][DR14]{2018ApJS..235...42A}, and the most recent release of the SDSS imaging data \citep[described in][DR13]{2017ApJS..233...25A}. Data of previous SDSS surveys are also included: the Legacy Spectra were finalized in DR8 \citep{2011ApJS..193...29A}, and the SEGUE-1 and SEGUE-2 spectra in DR9 \citep{2012ApJS..203...21A}. The MARVELS spectra were last re-reduced for DR12 \citep{2015ApJS..219...12A}.
\end{enumerate}

\begin{deluxetable*}{lll}
\tablecaption{New or Updated Value Added Catalogs\label{table:vac}}
\tablehead{\colhead{Description} & \colhead{Section} &  \colhead{Reference(s)}}
\startdata
{\bf Mini data release, 31 July 2018} \\
eBOSS DR14 QSO LSS catalogs & \S \ref{sec:eboss} &  \citet{ata18a} \\
eBOSS DR14 LRG LSS catalogs  & \S \ref{sec:eboss} &  \citet{bautista17b} \\
Optical Emission Line Properties and Black Hole Mass  & \S \ref{SPIDERS_VAC_lines} & \citet{Coffey2018}  \\
~~~~~~~~~~Estimates for SPIDERS DR14 Quasars \\
Open Cluster Chemical Abundance and Mapping Catalog & \S \ref{apogeevac} &  \citet{Donor2018} \\
{\bf DR15, 10 December 2018} \\
GEMA-VAC: Galaxy Environment for MaNGA VAC & \S \ref{manga:gema} & M. Argudo-Fernandez et al. (in prep). \\
MaNGA Spectroscopic Redshifts & \S \ref{manga:specz} &  \citet{2018MNRAS.477..195T} \\
MaNGA Pipe3D: Spatially resolved and integrated  &\S  \ref{mangaspec} & \citet{sanchez2016b,sanchez2018}  \\
~~~~~~~~~properties of DR15 MaNGA galaxies\footnote{update to DR14 VAC} \\
MaNGA Firefly Stellar Populations $^{\rm a}$ & \S \ref{mangaspec} & \citet{goddard2017,wilkinson2017,parikh2018} \\
MaNGA PyMorph DR15 photometric catalog & \S \ref{Mangamorph} &  \citet{Fischer2018} \\
MaNGA Morphology Deep Learning DR15 catalog & \S \ref{Mangamorph} & \citet{dominguezsanchez2018}\\
HI-MaNGA Data Release 1 & \S  \ref{himanga} &  \citet{Masters2018} \\
MaNGA Morphologies from Galaxy Zoo & \S \ref{Mangamorph} & \citet{GZ2,Hart16} 
\enddata
\end{deluxetable*}

\section{Data Distribution}
\label{sec:dataaccess}

The DR15 data can be accessed through a variety of mechanisms, depending on the type of data file and the needs of the user. All data access methods are described on the SDSS Web site (\url{https://www.sdss.org/dr15/data\_access}), and we also provide tutorials and examples for accessing and working with SDSS data products at (\url{https://www.sdss.org/dr15/tutorials}. We describe our four main data access mechanisms below.

All raw and processed imaging and spectroscopic data can be accessed through the Science Archive Server (SAS, {(\url{https://data.sdss.org/sas/dr15}). This site includes intermediate data products and VACs. The SAS is a file-based system, from which data can be directly downloaded by browsing, or in bulk mode using rsync, wget or Globus Online. Bulk downloading methods are outlined at \url{https://www.sdss.org/dr15/data\_access/bulk}. All data files available on the SAS have a data model (\url{https://data.sdss.org/datamodel}), which provides a detailed overview of the content of each data file.

Processed optical and infrared spectra, as well as processed imaging, can also be accessed on the SAS through the Science Archive Webapp (SAW), an interactive web application (webapp; \url{http://dr15.sdss.org} links to the DR15 version). In DR15, the SAW is serving MaStar spectra for the first time. Through this webapp, users can display individual spectra and overlay model fits included on the SAS. There is a search option available to select spectra based on e.g., plate number, coordinates, redshift or ancillary observing program. Searches can be saved for future references as permalinks. Spectra can be directly downloaded from the SAS through the webapp, and links are included to the SkyServer Explore page for each object. The user can select SDSS data releases back to DR8 (the SAW was originally developed during SDSS-III so serves data from that phase of SDSS onwards only), but is encouraged to always use the most recent data release at \url{https://data.sdss.org/home}. 

MaNGA datacubes and maps are not available in the SAW, but can be visualized and analyzed through Marvin. Marvin, described in detail in \S~\ref{sec:marvin}, provides links to the SAS for downloading data files, as well as the SkyServer Explore page.

The Catalog Archive Server \citep[CAS;][]{2008CSE....10....9T,2008CSE....10...30T} stores the catalogs of DR15: these include photometric, spectroscopic and derived properties. Some of the VACs also have catalogs that are stored on the CAS. The CAS can be accessed through the SkyServer webapp (\url{https://skyserver.sdss.org}), which provides Explore tools as well as the option of browser-based queries in synchronous mode. CASJobs (\url{https://skyserver.sdss.org/casjobs}) is suitable for more extensive queries, which are executed in asynchronous or batch mode, and offers users personal storage space for query results \citep{2008CSE....10...18L}. The CAS is integrated with SciServer (\url{https://www.sciserver.org}), which offers several data-driven science services, including SciServer Compute: a system that allows users to run Jupyter notebooks in Docker containers, directly accessing the SDSS catalogs.

All the data reduction software that is used by the various SDSS-IV teams to reduce and process their data (including links to the Marvin Repository on Github) is publicly available at \url{https://www.sdss.org/dr15/software/products}.

\section{MaNGA} 

\label{sec:manga}

The MaNGA survey uses a custom built instrument \citep{Drory2015} which feeds fibers from a suite of hexagonal bundles into the BOSS spectrograph \citep{Smee2013}. Over its planned five years of operations MaNGA aims to get data for $\sim$10,000 nearby galaxies (\citealt{Law2015,yan16,Yan2016}; and see \citet{Wake2017} for details on the sample selection).

DR15 consists of MaNGA observations taken during the first three years of the survey (up to Summer 2017) and nearly doubles the sample size of fully-reduced galaxy
data products previously released in DR14 \citep{2018ApJS..235...42A}.
These data products include raw data, intermediate reductions such as flux-calibrated spectra from individual exposures, and final data cubes and row-stacked
spectra (RSS) produced using the MaNGA Data Reduction Pipeline \citep[DRP;][]{law16}.
DR15 includes DRP data products for 4,824 MaNGA cubes distributed amongst 285 plates, corresponding to 4,621 unique galaxies plus 67 repeat observations and 118 special targets
from the ancillary programs (see \S \ref{sec:mangaancillary}).  Unlike in previous data releases, data cubes and summary RSS files are no longer produced for the twelve 7-fiber
mini-bundles on each plate that target bright stars and are used to derive the spectrophotometric calibration vector for each exposure \citep[see][]{yan16}; these observations will from here on instead be included
in the MaStar stellar spectral library (see \S \ref{sec:mastar}).

In addition, for the first time DR15 includes the release of derived spectroscopic products (e.g., stellar kinematics, emission-line diagnostic maps, etc.) from the MaNGA Data Analysis
Pipeline (K. Westfall et al., in prep; F. Belfiore et al., in prep), see \S \ref{sec:mangadap}.

We provide the sky footprint of MaNGA galaxies released in DR15 in Figure \ref{fig:mangasky}; while the projected final survey footprint is shown overlaid on the footprint of other relevant surveys and for two different expectations for weather at the telescope in Figure \ref{fig:mangaforecast}. 

\begin{figure*}
\centering
\includegraphics[angle=0,width=15cm]{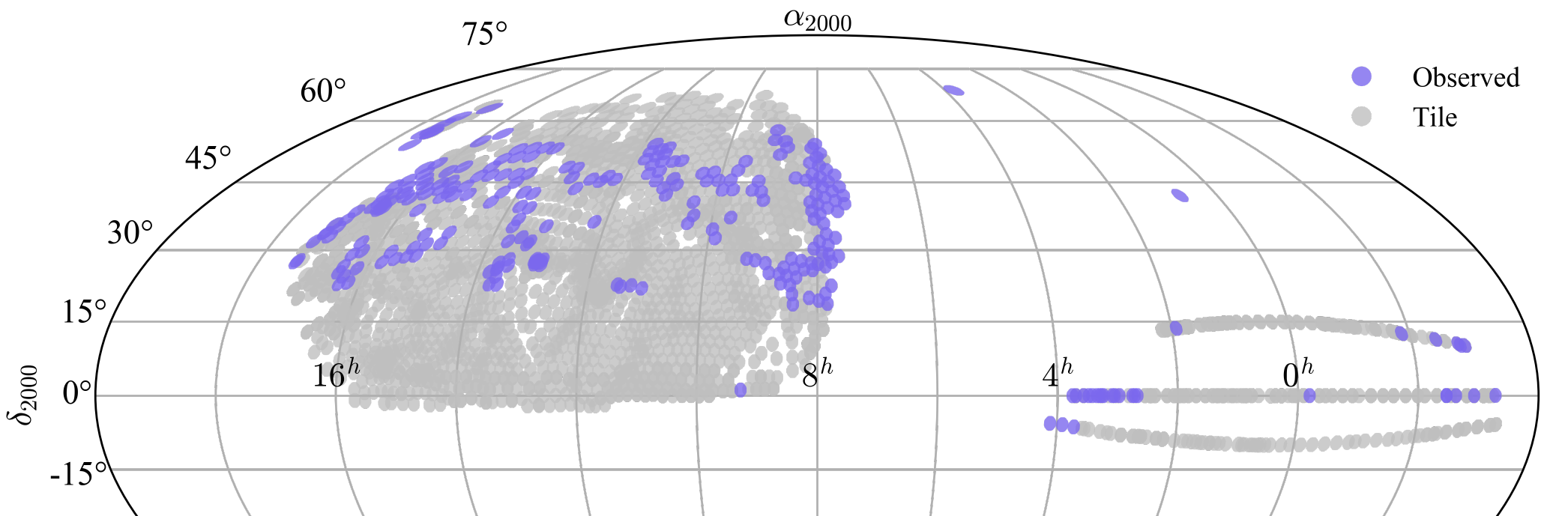}
\caption{The sky distribution (Mollewiede equatorial projection for Dec $>-20^\circ$) of MaNGA plates released in DR15 (purple). This is overlaid on a plot of all possible MaNGA plates (in grey). MaNGA targets are selected from a sample with SDSS-I photometry and redshifts; hence this footprint corresponds to the Data Release 7 imaging data \citep{DR7}.  Each plate contains 17 MaNGA targets, and around 30\% of all possible plates will be observed in the full 6-year survey. The most likely final footprint is indicated in Figure \ref{fig:mangaforecast}}
\label{fig:mangasky}
\end{figure*}

\begin{figure*}
\centering
\includegraphics[angle=0,width=15cm]{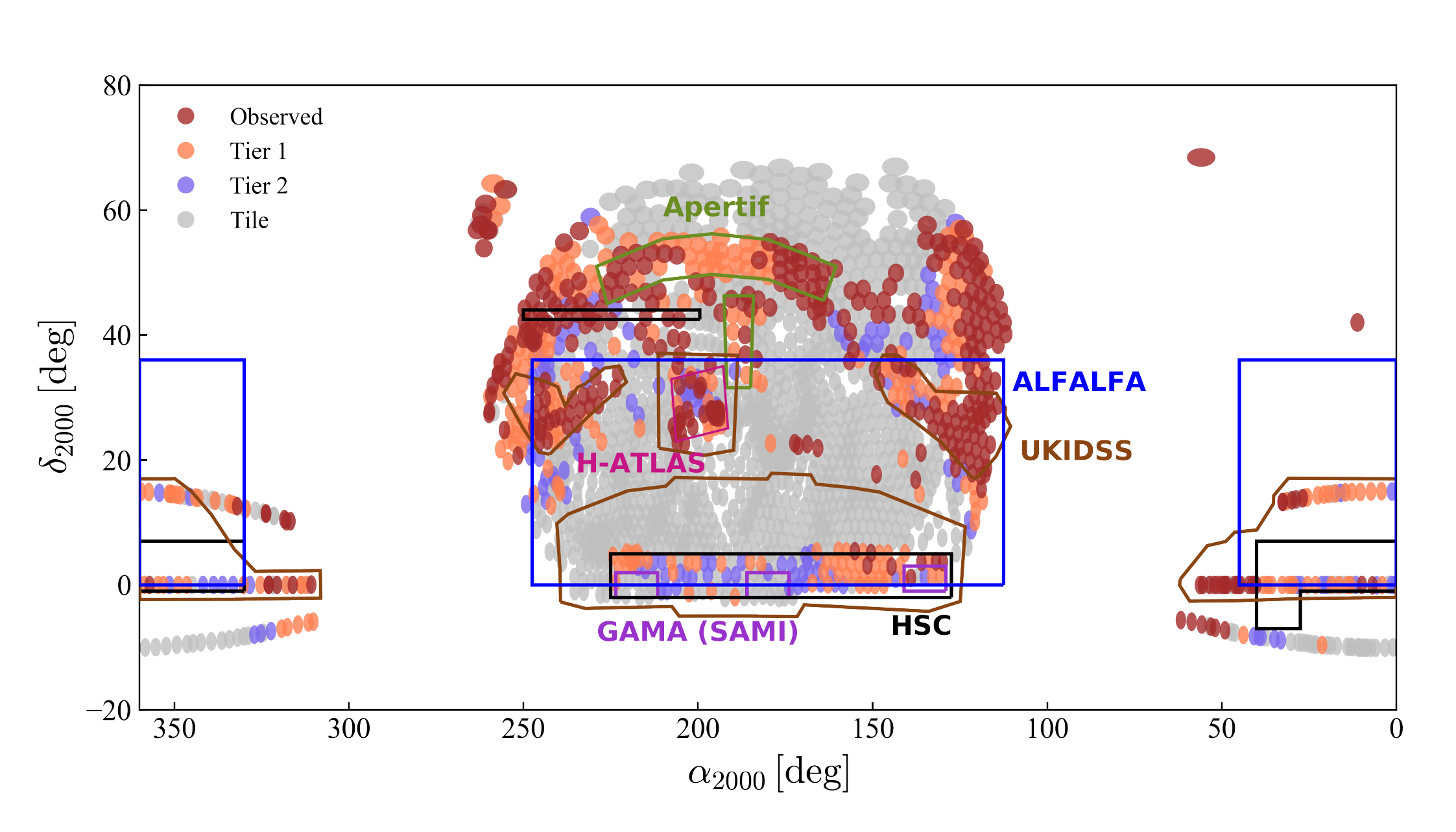}
\caption{The sky distribution (in a rectangular projection for clarity) of the MaNGA projected final footprint overlaid with information about other surveys. Because MaNGA targets are selected from a sample with SDSS-I photometry and redshifts, the selection of all possible plates (grey) corresponds to the Data Release 7 imaging data \citep{DR7}.  Each plate contains 17 MaNGA targets, and around 30\% of all possible plates will be observed in the full 6-year survey; this plot indicates the likely final footprint for (a) typical weather condition (Tier 1) and (b) good weather conditions (Tier 2). Completed plates noted on this plot show all observed plates at the time this was created; which is approximately one year of observing more than is being released in DR15. Where those plates are not filled in they have HI follow-up from the HI-MaNGA program (\citealt{Masters2018}; some, but not all of these data are released as a VAC in DR15 - see \S \ref{himanga}). For the most up-to-date version of this plot see \url{https://www.sdss.org/surveys/manga/forecast/}}
\label{fig:mangaforecast}
\end{figure*}

\subsection{MaNGA Data and Data Products}

\subsubsection{The Data Reduction Pipeline}
\label{sec:mangadrp}

The MaNGA Data Reduction Pipeline (DRP) is the IDL-based software suite that produces final flux-calibrated data cubes from the raw dispersed fiber spectra obtained at APO.  The DRP is described in detail
by \citet{law16} and consists of two stages.  The `2d' DRP processes individual exposures, applying bias and overscan corrections, extracting the one-dimensional fiber spectra,
sky-subtracting and flux-calibrating the spectra, and combining information from the four individual cameras in the BOSS spectrographs into a single set of row-stacked spectra ({\tt mgCFrame} files)
on a common wavelength grid.  The `3d' DRP uses astrometric information to combine the {\tt mgCFrame} fiber spectra from individual exposures into a composite data 
cube on a regularized 0.5$\arcsec$ grid, along with information
about the inverse variance, spaxel mask, instrumental resolution, and other key parameters.  The {\tt mgCFrame} per-exposure files are produced on both linear and logarithmic wavelength grids
directly from the raw detector pixel sampling, and used to construct the corresponding logarithmic and linearly-sampled data cubes.

The DRP data products release in DR15 are largely similar to those released in DR13 and DR14 (and identical to the internal collaboration release MPL-7), 
and consist of multi-extension FITS files giving the flux, inverse variance, mask, and other information for each object.  The metadata from all of our observations is summarized in a FITS binary table, 
``{\tt drpall-v2\_4\_3.fits}",  detailing the coordinates, targeting information, redshift, data quality, etc.
The version of the MaNGA DRP used for DR15 ({\tt v2\_4\_3}\footnote{\url{https://svn.sdss.org/public/repo/manga/mangadrp/tags/v2\_4\_3}}) incorporates some significant changes compared to the DR14 version
of the pipeline ({\tt v2\_1\_2}).  These changes include:

\begin{figure}
\centering
\includegraphics[angle=0,width=9cm]{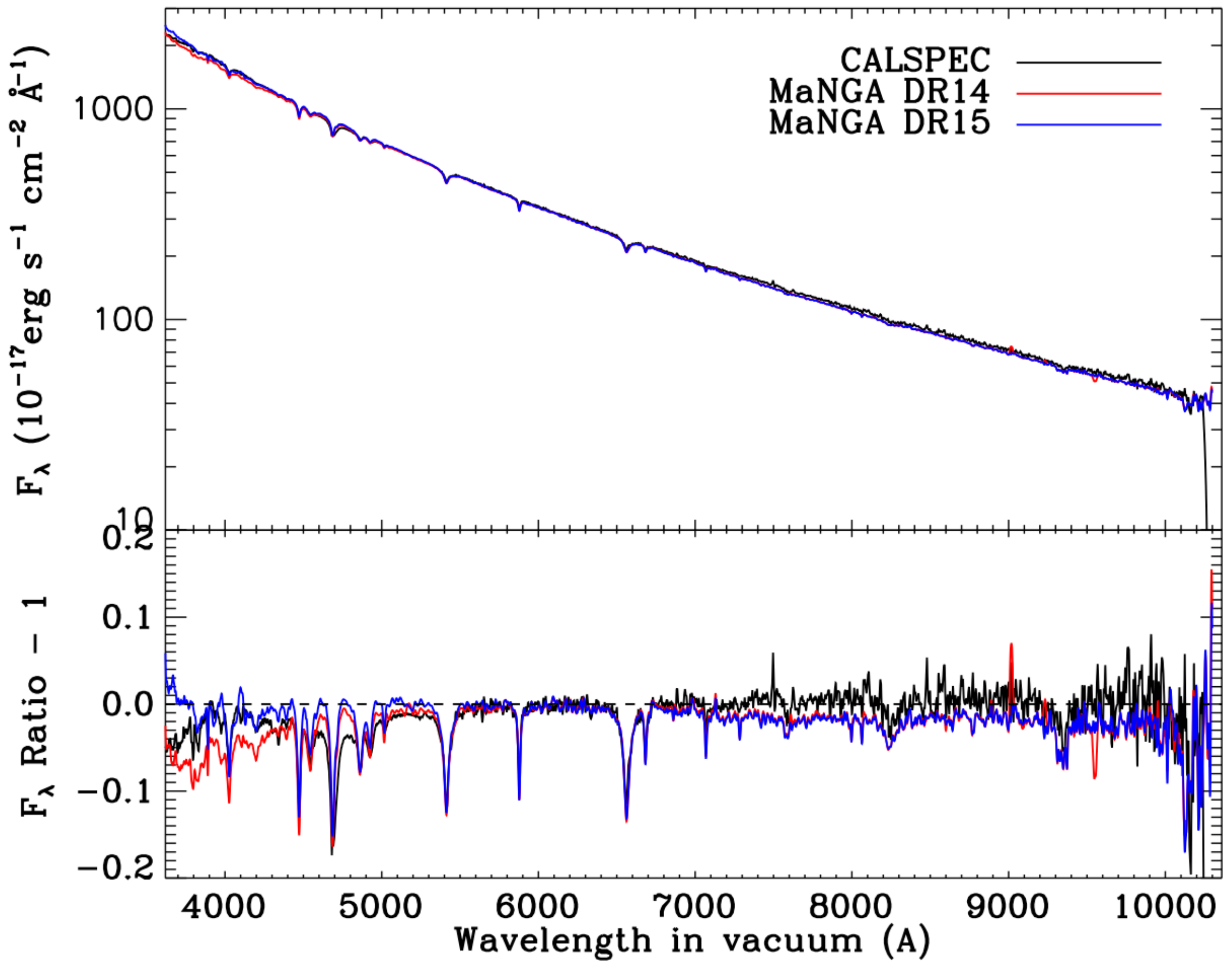}
\caption{This figure illustrates the flux calibration difference between DR14 and DR15 MaNGA data reductions. The upper panel shows the spectra for an Oke standard, HZ 21, a T=100,000~K star \citep{OkeShipman1971,Reynolds2003}, as given by the CALSPEC database (black) and by MaNGA in DR14 (red) and DR15 (blue) averaging over 9 exposures taken on plate 7444. The small difference at the blue wavelengths can be seen more obviously in the bottom panel where we divide these three spectra by a T=100,000~K blackbody spectrum normalized at 6000-6100A. Ignoring the absorption lines, this provides a test to our flux calibration. Using the BOSZ templates in DR15, the resulting continuum of this white dwarf agrees much better with the blackbody spectrum below 5000A, significantly improved compared to DR14, which uses the version of the Kurucz models \citep{Kurucz1979, Kurucz1981} which were produced in 2003, and to CALSPEC. (One can also compare this with Figure 9 of \citet{yan16}).
\label{fig:boszcalibration}}
\end{figure}

\begin{itemize}

\item The MaNGA DRP has been extended to produced one-dimensional reduced spectra for each of the MaStar targets observed during bright time; details of these modifications are described
in \S \ref{sec:mastar}.

\item  DR15 introduces some significant changes in the overall flux calibration relative to DR13/DR14 \citep[and relative to the description by][]{yan16}. Foremost among these is the use of BOSZ 
stellar spectral models \citep{bohlin17} instead of the original Kurucz templates built in 2003 to derive the spectrophotometric calibration based on contemporal observations of standard stars with the MaNGA 7-fiber mini-bundles. Since the BOSZ templates picked by the pipeline are bluer by 0.03 mag in SDSS $u-r$ color than the original Kurucz 2003 templates, this change slightly increases the overall flux blue-ward of 4000 \AA in the MaNGA data cubes. Test observations of hot white dwarfs compared to ideal blackbody models generally show better performance using the new BOSZ calibration (as illustrated in Figure \ref{fig:boszcalibration}). Additionally, the throughput loss vector applied to the observational data is now smoother at many wavelengths; high-frequency basis spline fits are still used in telluric regions, but the spline has a much lower frequency outside the telluric regions to avoid introducing artifacts due to slight template mismatches. This significantly reduces the amount of artificially high frequency low level ($\sim$ few percent) variations seen in the resulting spectra from earlier versions. The list of telluric regions is also updated. 

\item Many aspects of the spectral line-spread function (LSF) estimation in the DRP have changed in DR15 in order to improve the level of agreement with independent estimates 
(observations of bright stars and galaxies previously observed at higher spectral resolution, observations of the solar spectrum, etc.). These changes include the use of a Gaussian comb 
method to propagate LSF estimates through the wavelength rectification step, computation of both pre-pixelized and post-pixelized LSF estimates\footnote{{\it i.e.}, whether the best-fit Gaussian
model of the lines is determined by evaluation at the pixel midpoints (post-pixellized) or integrated over the pixel boundaries (pre-pixelized).  The two techniques can differ at the 10\% level
for marginally undersampled lines, and the appropriate value to use in later analyses depends on the fitting algorithm.}, improved interpolation over masked regions, 
and a modified arc lamp reference line list to improve LSF estimation and wavelength calibration in the far blue by rejecting poor-quality lines.
The DRP data products contain additional extensions to describe this new information, including a 3D cube describing the effective LSF at each spaxel within the MaNGA data cube as
a function of wavelength; this combines the information known about the LSF in each individual fiber spectrum to describe the net effect of stacking spectra with slightly different resolution.
The LSF changes and assessment against various observational calibrators will be described in greater detail by D. Law et al. (in prep).

\item  The DRP data cubes now contain extensions describing the spatial covariance introduced in the data cubes by the cube building algorithm.  This information is provided in the form of sparse
correlation matrices at the characteristic wavelengths of the SDSS $griz$ filters, and can be interpolated to estimate the correlation matrix at any other wavelength in the MaNGA data cubes. Note that the DR14 paper incorrectly stated that those data included these extensions. They did not (the team-internal MPL-5 which is the most similar MPL to DR14 did, but DR14 itself did not), so this is the first release of these extensions. 

\item The {\tt DRPall} summary file for DR15 contains ten additional columns with respect to DR14.  These columns include an estimate of the targeting redshift $z$ that is used as the starting guess
by the DAP when analyzing the MaNGA data cubes.  $z$ is generally identical to the NASA-Sloan Atlas (NSA) \citep{blanton2011} catalog redshift for the majority of MaNGA galaxies, but the origin of the redshift can vary for 
galaxies in the $\sim 25$ MaNGA ancillary programs.  Additional columns include a variety of estimates of the volume weights for the MaNGA primary and secondary galaxy samples.

\item Additional under-the-hood modifications to the DRP have been made for DR15 that provide minor bug fixes and performance improvements.  These include modifications to the reference pixel
flat-fields for certain MJDs, updates to the reference bias and bad pixel masks, better rejection of saturated pixels, 
updates to the algorithms governing weighting of the wavelength rectification algorithm near ultra-bright emission lines, etc.
A detailed change-log can be found in the DRP online repository.\footnote{\url{https://svn.sdss.org/public/repo/manga/mangadrp/tags/v2\_4\_3/RELEASE\_NOTES}}

\end{itemize}

When working with the MaNGA data, note that there are several quality-control features that should be used to ensure the best scientific quality output.
First, each MaNGA data cube has a FITS header keyword {\tt DRP3QUAL} that describes the overall quality of the cube (identifying issues such as focus problems, flux calibration problems,
large numbers of dead fibers, etc.).  About 1\% of the data cubes are flagged as significantly problematic (i.e., have the CRITICAL quality bit set) and should be treated with extreme caution.
Additionally, there is a 3d mask extension to each data cube that contains spaxel-by-spaxel information about problematic regions within the cube.  This mask identifies issues such as dead fibers
(which can cause local glitches and holes within the cube), foreground stars that should be masked by analysis packages such as the DAP, etc.
Although the vast majority of cosmic rays and other transient features are detected by the DRP and flagged (either for removal or masking), lower-intensity glitches (e.g, where the edge of a cosmic ray
track intersects with a bright emission line) can sometimes be missed and propagate into the final datacubes where they show up as unmasked hot pixels.  
Future improvements to the DRP may further address this issue, but caution is thus always advised
when searching for isolated emission features in the data cubes.

For information on downloading MaNGA data in DR15 please see \S \ref{sec:dataaccess}; new for DR15 is the Marvin interface to MaNGA data (see \S \ref{sec:marvin} below). 

\subsubsection{The Data Analysis Pipeline}

\label{sec:mangadap}

The MaNGA data-analysis pipeline (DAP) is the SDSS-IV software
package that analyzes the data produced by the MaNGA data-reduction
pipeline (DRP).  The DAP currently focuses on ``model-independent''
properties; i.e., those relatively basic spectral properties that
require minimal assumptions to derive.  For DR15, these products include
stellar and ionized-gas kinematics, nebular emission-line fluxes and
equivalent widths, and spectral indices for numerous absorption
features, such as the Lick indices \citep{1997ApJS..111..377W,
1998ApJS..116....1T} and D4000 \citep{1983ApJ...273..105B}.  Examples of
the DAP provided measurements and model fits are shown in Figure
\ref{fig:dap_example}, discussed through the rest of this Section.

\begin{figure*}[htp]
\begin{center}
\includegraphics[width=0.8\textwidth]{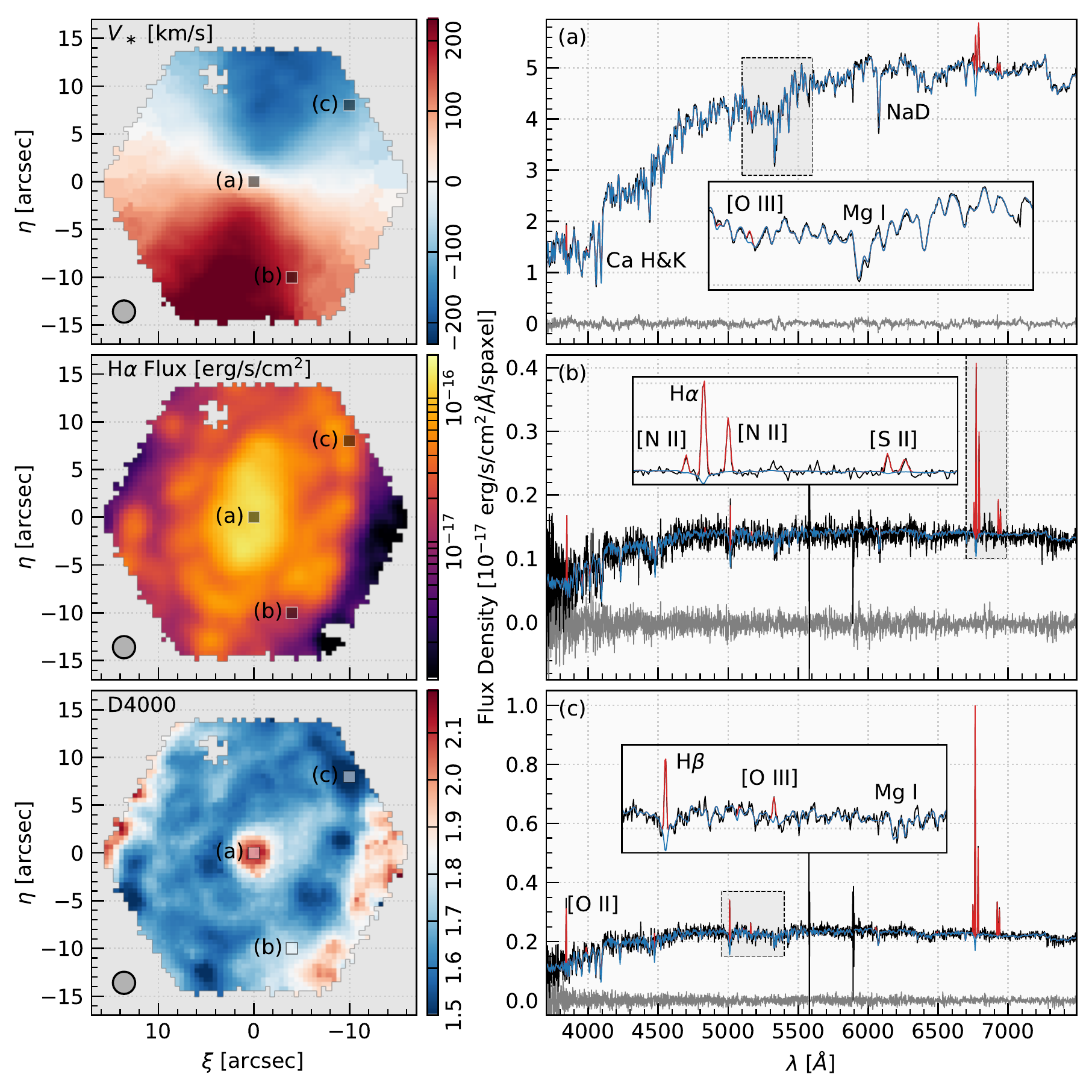}
\end{center}
\caption{
Example data provide by the MaNGA data-analysis pipeline (DAP) for MaNGA
observation 8138-12704, MaNGA ID 1-339041, following the hybrid binning
approach ({\tt DAPTYPE} is {\tt HYB10-GAU-MILESHC}).  The left columns
shows maps, or images, of some of the DAP derived quantities.  Namely,
from top to bottom, the stellar velocity field, H$\alpha$ flux, and
D4000 spectral index, where the measured value is indicated by the
colorbar to the right of each map panel.  The effective beam size for
the MaNGA observations (FWHM$\sim$$2\farcs5$) is shown by the gray
circle in the bottom left of each map panel.  Three spaxels are
highlighted and labeled as {\it (a)}, {\it (b)}, and {\it (c)},
according to their spectra plotted in the right column.  Each spectrum
panel shows the observed MaNGA spectrum ({\it black}),
stellar-continuum-only model ({\it blue}), and best-fitting
(stars$+$emission lines) model ({\it red}); the residuals between the
data ({\it black}) and the model ({\it red}) are shown in gray.  Note
that the red and blue lines are identical except for regions with
nebular emission.  A few salient emission and absorption features are
marked in each panel.  Inset panels provide a more detailed view of the
quality of the fitted models in the regions highlighted with gray boxes.
The spectrum panels only show the spectral regions fit by the DAP, which
is limited by the MILES spectral range for DR15.
}
\label{fig:dap_example}
\end{figure*}

An overview of the DAP is provided by K.~Westfall et al., (in prep).
There, we describe the general workflow of the pipeline, explain the
detailed algorithm used for each of its primary products, provide
high-level assessments of its performance, and describe the delivered
data products in detail.  In-depth assessments of the stellar
kinematics, ionized-gas kinematics, emission-line fluxes, and
emission-line equivalent widths are provided by K.~Westfall et al., 
(in prep), and F.~Belfiore et al., (in prep). 
All survey-provided properties are currently derived from the datacubes
sampled in constant steps of the logarithm of wavelength (i.e., the {\tt LOGCUBE} files).
However, the core functions are developed to consider each spectrum
largely independently.

The DAP allows for a number of different options when analyzing the
data, which we refer to as the analysis mode or {\tt DAPTYPE}.  In DR15,
the {\tt DAPTYPE} joins the keywords identifying the type of spatial
binning applied (e.g., Voronoi binned to S/N$\gtrsim10$, {\tt VOR10}),
the parametric form of the line-of-sight velocity distribution (LOSVD)
used for the stellar kinematics (a Gaussian function, {\tt GAU}), and
the template set used to model the stellar continuum (a hierarchically
clustered distillation of the MILES stellar library, {\tt MILESHC}).
For DR15, two {\tt DAPTYPE}s have been made available, {\tt
VOR10-GAU-MILESHC} and {\tt HYB10-GAU-MILESHC}.  That is, only the
binning approach differs between the two available {\tt DAPTYPE}s,
primarily distinguishing whether or not the main analysis steps are
performed on binned spectra or individual spaxels.  The stellar LOSVD is
always assumed to be Gaussian, and the 42 templates resulting from a
hierarchical clustering analysis of the MILES stellar library
\citep{MILES,2011A&A...532A..95F} are always used for the continuum
templates; details regarding the latter is discussed in K.~Westfall et
al., (in prep).

In the first mode ({\tt VOR10-GAU-MILESHC}), the spaxels are binned
using the Voronoi-binning scheme from \citet{2003MNRAS.342..345C} to a
minimum $g$-band S/N of 10 per spectral pixel.  The first mode then
performs {\it all} subsequent analysis on those binned spectra.
Alternatively, the second mode ({\tt HYB10-GAU-MILESHC}) only performs
the stellar kinematics on the binned spectra; the subsequent
emission-line and spectral-index measurements are all performed on the
individual spaxels.  This ``hybrid'' binning approach is likely the
approach that most users will want to use in their analysis.  The main
exception to this is if any subsequent analyses depend on, e.g., the
availability of emission-line models for the binned spectra, as is the
case for the {\tt FIREFLY} VAC \citep[see \S
\ref{mangaspec}]{wilkinson2017}.  The example data shown in Figure
\ref{fig:dap_example} is for observation 8138-12704 following from the
hybrid binning approach.  Close inspection of the stellar velocity field
will show that outermost regions have been binned together, all showing
the same stellar velocity measurement.  However, the H$\alpha$ flux and
D4000 maps have measurements for each spaxel.

The DAP is executed for all observations obtained by the MaNGA survey;
however, some observations, primarily those obtained for our ancillary
science programs, do not have all the required parameters currently
needed as input by the DAP.  Additionally, a few observations ($<$0.3\%)
trip corner failure modes of the DAP leading to errors in the
construction of its main output files.  These issues mean that not all
{\tt LOGCUBE} files provided by the DRP have associated DAP products.
For those observations that are successfully analyzed (4718 in total),
the DAP provides two main output files for each {\tt DAPTYPE}, the {\tt
MAPS} file and the model {\tt LOGCUBE} file.  Examples of how to access
and plot the data in these files are provided in a set of tutorials on
the data-release website at \url{https://www.sdss.org/dr15/manga/manga-tutorials/dap-tutorial/}.

The {\tt MAPS} file contains all of the derived properties organized as
a series of maps, or images, that have the same on-sky projection as a
single wavelength channel in the analyzed DRP {\tt LOGCUBE} file.  The
images in the left panels of Figure \ref{fig:dap_example} are example
maps taken from the DAP {\tt MAPS} file for observation 8138-12704.  The
maps are organized in a series of extensions grouped by the measurement
they provide.  Some extensions contain a single image with all of the
relevant data, whereas other extensions have multiple images, one for,
e.g., each of the measured emission lines.  For example, the {\tt
STELLAR\_VEL} extension has a single image with the measured
single-component stellar velocity measured for each spatial bin (like
that shown in Figure \ref{fig:dap_example}), while the {\tt SPECINDEX}
extension has 46 images, organized similarly to the wavelength channels
in the DRP datacubes (the D4000 map shown in Figure
\ref{fig:dap_example} is in the 44th channel of the {\tt SPECINDEX}
extension).  

The DAP-output model {\tt LOGCUBE} file provides both the S/N-binned
spectra and the best-fitting model spectra.  From these files, users can
plot the best-fitting model spectra against the data, as demonstrated in
Figure \ref{fig:dap_example}, as an assessment of the success of the
DAP.  This is particularly useful when a result of the fit, e.g. the
H$\alpha$ flux, seems questionable.  Indeed, K.~Westfall et al., {\it in
prep}, note regimes where the DAP has not been appropriately tailored to
provide a successful fit; this is particularly true for spectra with
very broad emission lines, such as the broad-line regions of AGN.  Users
are encouraged to make sure they are well aware of these limitations in
the context of their science goals.  Finally, in combination with the
DRP {\tt LOGCUBE} file, users can use the model {\tt LOGCUBE} data to
construct emission-line-only or stellar-continuum-only data cubes by
subtracting the relevant model data.


Although we have endeavored to make the output data user-friendly, there
are a few usage quirks of which users should be aware:
\begin{enumerate}
\item As with all SDSS data products, users are strongly encouraged to
understand and use the provided quality flags, for these data provided
as masks.  The mask bits provide important information as to whether or
not users should trust the provided measurements in their particular use
case.  The conservative approach of ignoring any measurement where the
mask bit is nonzero is safe, at least in the sense of not including any
measurements we {\it know} to be dubious.  However, the DAP makes use of
an {\tt UNRELIABLE} flag that is intended to be more of a warning that
users should consider how the measurements affect their science as
opposed to an outright rejection of the value.  The {\tt UNRELIABLE}
flag is put to limited use in DR15, only flagging measurements that
hinge on bandpass integrals (emission-line moments and non-parametric
equivalent widths, and spectral indices) where any pixels are masked
within the bandpass.  However, this bit may become more extensively used
in future releases as we continue to vet the results of the analysis.  A
more extensive discussion of the mask bits and their usage is provided
by K.~Westfall et al., (in prep).
\item To keep the format of the output files consistent with the DRP
{\tt LOGCUBE} files, the binned spectra and binned-spectra measurements
are repeated for each spaxel within a given bin.  This means that, e.g.,
the stellar velocity dispersion measured for a given binned spectrum is
provided in the output DAP map at the location of each spaxel in that
bin.  Of course, when analyzing the output, one should most often only
be concerned with the {\it unique} measurements for each observation.
To this end, we provide an extension in the {\tt MAPS} file that
provides a ``bin ID'' for each spaxel.  Spaxels excluded from any
analysis (as in the buffer region during the datacube construction) are
given a bin ID of -1.  This allows the user to select all the unique
measurements by finding the locations of all unique bin ID values,
ignoring anything with a bin ID of -1.  Tutorials for selecting the
unique measurements in the DAP output maps are provided via the
data-release website at
\url{https://www.sdss.org/dr15/manga/manga-tutorials/dap-tutorial/}.
\item Corrections are provided for a few quantities in the {\tt MAPS}
file {\it that have not been applied to the data in the output files}.
The stellar velocity dispersion and ionized-gas velocity dispersions are
provided {\it as measured} from the core {\tt pPXF} software
\citep{2004PASP..116..138C, 2017MNRAS.466..798C} used by the DAP.  This
means that any instrumental effects present during the fitting process
are also present in the output data.  For both the stellar and
ionized-gas dispersions, we have estimated the instrumental corrections
for each measurement and provided the result in extensions in the {\tt
MAPS} file.  {\it These corrections should be applied when using the
data for science.}  For the velocity dispersion measurements, our
purpose in not applying the corrections ourselves is to allow the user
freedom in how they deal with measurements of the dispersion that are
{\it below} our measurement of the instrumental resolution.  Such issues
can be pernicious at low velocity dispersion and the treatment of these
data can have significant effects on, e.g., the construction of a
radially averaged velocity dispersion profile (see K.~Westfall et al.,
in prep. who discuss this at length, and also \citealt{Penny2016}
who discuss this issue for dwarf galaxies).  Corrections are also
provided (but {\it not} applied) for the spectral indices to convert the
measurement to zero velocity dispersion at the spectral resolution of
the MILES stellar templates \citep{2011A&A...531A.109B} used during the stellar-continuum fit.
Additional detail regarding these corrections is provided in K. Westfall et
al., (in prep.) and tutorials demonstrating how to apply them to the data are
provided via the data-release website.
\item In the hybrid binning scheme, the stellar kinematics are performed
on the binned spectra, but the emission-line fits are performed on the
individual spaxels.  When comparing the model to the data, the user must
compare the emission-line modeling results to the DRP {\tt LOGCUBE}
spectra, {\it not} the binned spectra provided in the DAP model {\tt
LOGCUBE} file, unless the ``binned'' spectrum is actually from a single
spaxel.  Tutorials for how to overplot the correct stellar-continuum and
emission-line models are provided via the data-release website.
\end{enumerate}


Finally, similar to the {\tt DRPall} file provided by the MaNGA DRP, the
DAP constructs a summary table called the {\tt DAPall} file.  This
summary file collates useful data from the output DAP files, as well as
providing some global quantities drawn from basic assessments of the
output maps, that may be useful for sample selection.  For example, the
{\tt DAPall} file provides the luminosity-weighted stellar velocity
dispersion and integrated star-formation rate within 1 $R_e$.  The
sophistication of these measurements are limited in some cases.  For
example, the star-formation rate provided is simply based on the
measured H$\alpha$ luminosity and does not account for internal
attenuation or sources of H$\alpha$ emission that are unrelated to star
formation, as such we caution users to make use of this for science only
after understanding the implications of this caveat.  Development and
refinement of {\tt DAPall} output will continue based on internal and
community input.  Additional discussion of how these properties are
derived is provided by K.~Westfall et al., (in prep).

\subsection{Marvin Access to MaNGA}
\label{sec:marvin}

{\tt Marvin} \citep{brian_cherinka_2018_1230529}\footnote{\url{https://www.sdss.org/dr15/manga/marvin/}}  is a new tool designed for streamlined access to the MaNGA data, optimized for overcoming the challenges of searching, accessing, and visualizing the complexity of the MaNGA dataset. Whereas previous generations of SDSS took spectra only of the centers of galaxies, MaNGA takes many spectra of each galaxy, in a hexagonal grid across the face of each (IFU bundle), which are combined into a final data cube. This means that for each object there is not a single spectrum, but in fact a suite of complex results in one or more data cubes. The motivation of {\tt Marvin} arises from the additional complexity of MaNGA data, namely, the spatial interconnectivity of its spectra.

{\tt Marvin} allows the user to: 
\begin{itemize}
\item access reduced MaNGA datacubes local, remotely, or via a web interface.
\item access and visualize data analysis products.
\item perform powerful queries on metadata.
\item abstract the MaNGA datamodel and write code which is agnostic to where the data actually lives.
\item make better visualization and scientific decisions by mitigating common mistakes when accessing these type of data.
\end{itemize}

{\tt Marvin} has two main components: a webapp and a Python package of tools, both using an underlying {\tt Marvin} API (or Application Programming Interface). The webapp, {\tt Marvin Web}\footnote{\url{https://dr15.sdss.org/marvin}}, provides an easily accessible interface for searching the MaNGA dataset and visual exploration of individual MaNGA galaxies. The {\tt Marvin} suite of Python tools,  {\tt Marvin Tools}, provides seamless programmatic access to the MaNGA data for more in-depth scientific analysis and inclusion in your science workflow. {\tt Marvin} contains a multi-modal data access system that provides remote access to MaNGA files or sub-data contained within, download MaNGA files to work with on the users local machine, and seamlessly transition between the two with a negligible change in syntax.

Existing 3d data cube visualizers in astronomy, as well as in other scientific disciplines, often come as standalone desktop applications designed to visualize and interact with individual files local to a client machine.  However, these tools are highly specific, limited to exploring files one at a time, and still require manually downloading all data locally. While {\tt Marvin} is a tool for 3d cube visualization, its focus is on streamlined data access from local or remote sources, with a clear separation of components into browser-based visualization and programmatic data tools, rather than on providing yet another desktop-based cube viewer.  {\tt Marvin}'s design allows for users to rapidly explore and access the data in a manner of their choosing, whilst still providing enough flexibility to, if desired, plug the data into existing cube viewers available in the astronomy community.

The components of {\tt Marvin} are described in more detail in the {\tt Marvin} paper (B. Cherinka et al. in prep.) as well as in the {\tt Marvin} documentation\footnote{\url{https://sdss-marvin.readthedocs.io/en/stable/}}, which also contains tutorials and example Jupyter notebooks. In addition, we briefly introduce them below.

\subsubsection{Marvin Web}

The {\tt Marvin Web} provides quick visual access to the set of MaNGA galaxies. It provides a dynamic, interactive, point-and-click view of individual galaxies to explore the output from the MaNGA DRP and DAP (\S \ref{sec:mangadrp} \S \ref{sec:mangadap} respectively), along with galaxy information from the NSA catalog \citep{blanton2011} \footnote{\url{https://www.sdss.org/dr15/manga/manga-target-selection/nsa/}}.

We show a screen-shot of the View-Spectra page of {\tt Marvin Web} in Figure~\ref{Fig:MarvinWebMaps}. By clicking anywhere within the galaxy IFU bundle on the SDSS three-color image, or any Data Analysis 2D Map, the user can explore the spectrum at that location for quick inspection. The visualized spectrum is interactive as well, allowing panning and zooming.

Additional pages {\tt Marvin Web} provides are:
\begin{itemize}
\item a Query page, for searching the MaNGA dataset through an SQL-like interface.
\item a Plate page, containing all MaNGA galaxies observed on a given SDSS plate.
\item and an Image Roulette page, for randomly sampling images of MaNGA galaxies.
\end{itemize}

Tutorials for navigating {\tt Marvin Web} can be found at \url{https://www.sdss.org/dr15/manga/manga-tutorials/marvin-tutorial/marvin-web/}.

{\tt Marvin Web} is designed as a gateway to entry into real MaNGA data, providing commonly-desired functionality all in one location, as well as code snippets to help transition users into a more programmatic environment using the {\tt Marvin Tools}.

\begin{figure*}
\centering
\includegraphics[width=2.\columnwidth]{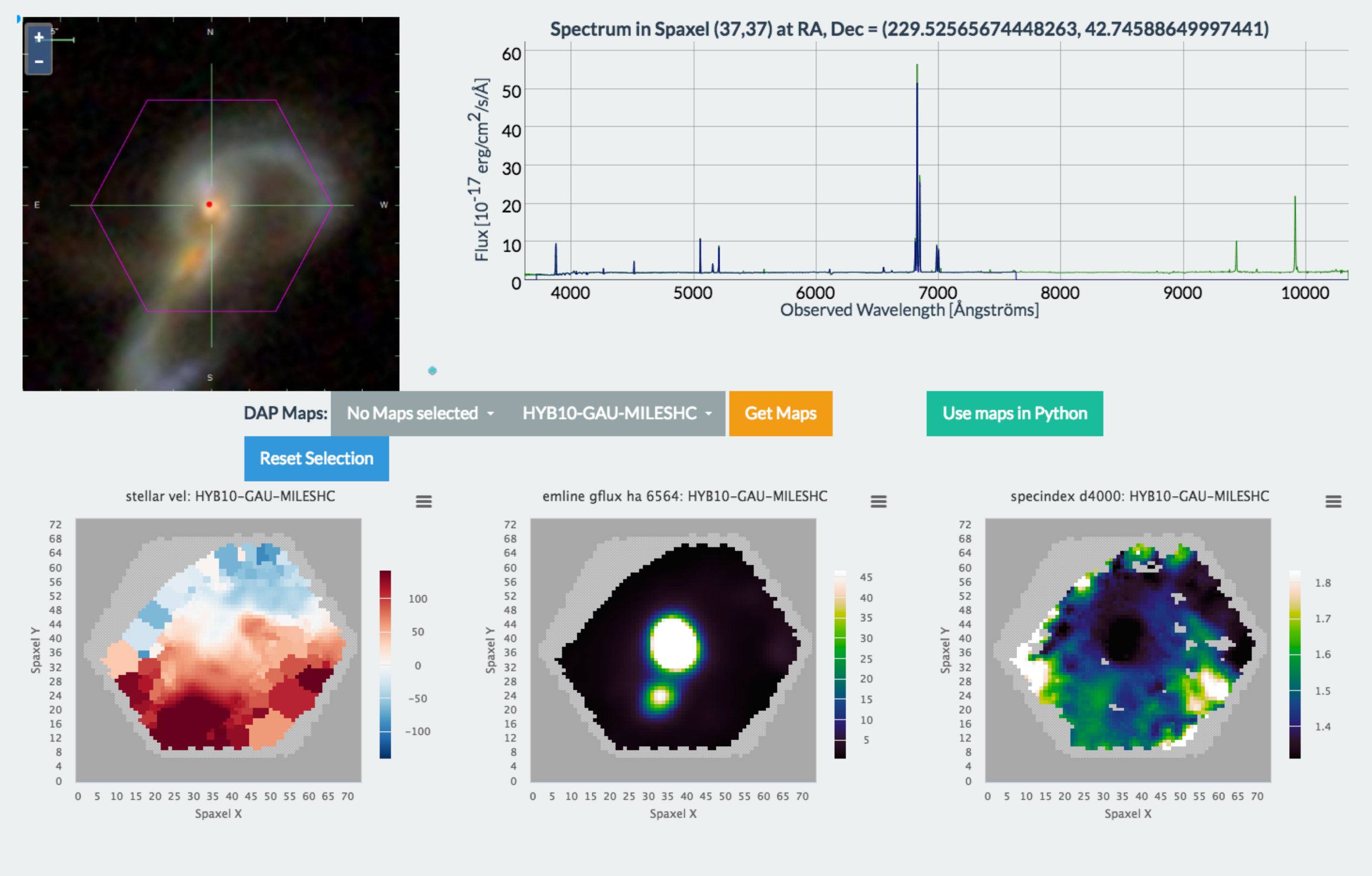}
\caption{Screenshot of the galaxy maps view of the {\tt Marvin Web} for the MaNGA galaxy 12-193481 (Mrk 848). The SDSS three-color image of the galaxy is shown in the top left part of the figure. The upper right panel shows the spectrum of the spaxel at the position (37,37), which corresponds to the center of the bundle (shown by the red dot). The maps show: (lower left) stellar kinematics; (lower middle) H$\alpha$ emission line flux; and (lower right) D4000 spectral index for this galaxy based on its unbinned spectral data cube from the MaNGA DAP (see \S \ref{sec:mangadap}).}
\label{Fig:MarvinWebMaps}
\end{figure*}

\subsubsection{The Marvin Tools}

{\tt Marvin Tools} provides a programmatic interaction with the MaNGA data, enabling rigorous and repeatable science-grade analyses. {\tt Marvin Tools} come in the form of a Python package that provides convenience classes and functions that simplify the processes of searching, accessing, downloading, and interacting with MaNGA data, selecting a sample, running user-defined analysis code, and producing publication quality figures.

The {\tt Marvin Tools} are a pip-installable product, packaged under \texttt{sdss-marvin}, with full installation instructions at the Marvin documentation\footnote{\url{https://sdss-marvin.readthedocs.io/en/stable/installation.html}}, and source code on Github\footnote{\url{https://github.com/sdss/marvin}}.

Overall, {\tt Marvin Tools} allow for easier access to the data without knowing much about the data model, by seamlessly connecting all the MaNGA data products, eliminating the need to micromanage a multitude of files.  The user can do all their analysis from one interface.

\subsubsection{Queries in Marvin}

Both {\tt Marvin Web} and {\tt Tools} provide interfaces for searching the MaNGA dataset through a Structured Query Language ({\tt SQL})-like interface, either via a web-form or a Python class. The {\tt Marvin} Query system uses a simplified {\tt SQL} syntax that focuses only on a filter condition using boolean logic operators, and a list of parameters to return. This eliminates the need to learn the full SQL language and the detailed MaNGA database layout. With this query system, users can make queries across the entire MaNGA sample using traditional global galaxy properties (functionality to perform intra-galaxy queries using individual spaxel measurements is planned for a future release). Tutorials for querying with {\tt Marvin} can be found for the web\footnote{\url{https://www.sdss.org/dr15/manga/manga-tutorials/marvin-tutorial/marvin-web/}} and for the tools\footnote{\url{https://sdss- marvin.readthedocs.io/en/stable/query.html}}.

\subsection{MaStar: A Large and Comprehensive Stellar Spectral Library}

\label{sec:mastar}

Stellar spectral libraries are an essential tool for many fields in astronomy. They are especially useful for modeling spectra of external galaxies, including fitting for redshift and stellar kinematics, fitting the continuum to isolate emission lines, and calculating stellar population models \citep[e.g.][]{Leitherer1999, BruzualCharlot2003, Maraston2005, Vazdekis2010, Conroy2010, Conroy2013} for deriving age, metallicity, and stellar mass of the stellar populations from integrated light spectra. They are also useful for Galactic astronomy and stellar astronomy. Although theoretical spectral libraries have been substantially improved over the years, they are still not realistic enough for certain stellar types (e.g. very cold stars and Carbon stars) due to the incomplete line list and difficult-to-model physical effects, such as convection, microturbulence, and deviations from plane-parallel geometry and local thermodynamic equilibrium (non-LTE). Therefore, empirical libraries are still needed for many applications and for calibrating the theoretical models, provided one is able to assign robust stellar parameters to the empirical spectra. 

At the beginning of the MaNGA survey, there were no empirical stellar libraries available covering the whole MaNGA wavelength range with a spectral resolution that is equal to or higher than that of MaNGA. Current state-of-the-art empirical stellar libraries also have some other shortcomings. Some libraries have issues with flux calibration or telluric subtraction. Furthermore, all existing libraries have limited stellar parameter space coverage, lacking sufficient sampling in especially cool dwarfs, carbon stars, metal-poor stars, and very hot stars. They also do not sufficiently sample the [${\rm \alpha}$/Fe] vs. [Fe/H] space (see \citealt{Maraston2011}) for a discussion of all these problems). These issues prompted us to take advantage of a parallel observing opportunity in SDSS-IV for assembling an empirical stellar spectral library that samples a wider stellar parameter space with a larger number of stars than any previous library, and matches MaNGA's wavelength coverage and spectral resolution.

Included in this data release is the first version of the MaNGA Stellar Library (MaStar). These observations are performed by piggybacking on the APOGEE-2N observations during bright time. MaNGA fiber bundles are plugged along with APOGEE fibers on these APOGEE-led plates to observe selected stars. As a result, the MaStar stellar spectra are observed using exactly the same instrument as MaNGA galaxies so they provide an ideal set of templates for modeling stellar continuum and stellar populations in MaNGA galaxies.

The program has so far observed several thousands of stars, each with several epochs of observation. The version we are releasing in DR15 includes 8646 good quality spectra for 3321 unique stars, which cover a wide range in stellar parameter space. The details of the target selection, data reduction, flux calibration, and stellar parameter distribution are described by \citet{Yan2018}. Here we provide a brief summary. 

\subsubsection{Target Selection}

A good target selection is essential for achieving a wide sampling of stellar parameter space. We aim to cover the stellar parameter space as completely as possible and sample it roughly evenly. We base our selection primarily on existing stellar parameter catalogs, including APOGEE-1 and -2 \citep{Majewski2017}, SEGUE \citep{2009AJ....137.4377Y}, and LAMOST \citep{LAMOST}. Given the field plan of APOGEE-2, we select all the stars available in these catalogs in the planned APOGEE-2 footprint. For each star, we count its neighboring stars in stellar parameter space ($T_{\rm eff}$, $\log g$, [Fe/H]) and assign it a selection weight that is inversely proportional to its number of neighbors. The number of APOGEE-2 targeting designs for each field is also taken into account. We then draw our targets randomly in proportion to the normalized selection weight. This method flattens the stellar parameter space distribution and picks rare stars in those fields where they are available. 

In fields without stars with known stellar parameters, we use spectral energy distribution (SED) fitting to search for hot and cool stars to patch the stellar parameter distribution at the hot and cool ends. 

The targets are required to have $g$ or $i$-band magnitude brighter than 17.5\ in order to achieve a signal-to-noise greater than 50 per pixel in 3 hours of integration, although not all fields have the same integration time or the same number of visits. They are also required to be fainter than 12.7 magnitude in both $g$ and $i$-band in order to stay below the saturation limit of the detector for 15 minute exposures. We later lowered the saturation limit to 11.7 to include more luminous stars, with a slight offset in fiber placement for stars with magnitudes between 11.7 and 12.7.  This slight offset does not affect our flux calibration due to our unique calibration procedure. 

These magnitude limits yield relatively few OB stars and blue supergiants, as they have to be very distant or very extincted to fall within this magnitude range. Therefore, currently we are adjusting our exposure time in certain fields to expand our parameter space distribution in the blue and luminous end. The first version of the library does not have many such stars but we will improved on this for the final version which we expect to come out in the final SDSS-IV Data Release. 

\subsubsection{Observations}

Observations for MaStar are obtained in a similar fashion to the MaNGA observations except that they are conducted under bright time and without dithering. Since we are piggybacking on APOGEE-2, if APOGEE-2 visits a field multiple times, we would obtain multiple visits for the stars on that plate as well. Therefore, some stars have many visits and some stars have only 1 visit. Each visit of APOGEE-2 is typically 67 minutes long, which would allow us to take four 15-minutes exposures, unless interrupted by weather or other reasons. Each plate has 17 science targets and 12 standard stars, same as MaNGA. We take flat and arc frames before each visit. 

\subsubsection{Data Reduction}

The reduction of the MaStar data is handled by the MaNGA Data Reduction Pipeline (DRP; see \S \ref{sec:mangadrp}, \citealt{law16}). It has two stages. The first stage processes the raw calibration frames and science frames to produce the sky-subtracted, flux-calibrated, camera-combined spectra for each fiber in each exposure. The second stage differs between MaNGA galaxy data and MaStar stellar data. For MaStar stellar data, we evaluate the flux ratios among fibers in a bundle as a function of wavelength, and constrain the exact location of the star relative to the fiber positions. This procedure helps us derive the light loss due to the finite fiber aperture as a function of wavelength. The procedure takes into account the profile of the PSF and the differential atmosphere refraction. It is similar to how we handle flux calibration in MaNGA data \citep{yan16}. We then correct the spectra for this aperture-induced light loss and arrive at the final flux-calibrated stellar spectra. Comparison with photometry shows that our relative flux calibration are accurate to 5\% between $g$ and $r$ -band, and to 3\% between $r$ and $i$, and between $i$ and $z$  bands.

For each star, we combine the spectra from multiple exposures on the same night, and refer to these combined spectra as ``visit spectra". We do not combine spectra from different nights together for the same star because they can have different instrumental resolution vectors and some stars could be variable stars.  By summer 2017, we have obtained 17,309 visit-spectra for 6,042 unique stars. Because not all visit-spectra are of high quality, as we will discuss below, we selected only those with high quality and present this subset as the primary set to be released. The primary set contains 8646 visit spectra for 3321 unique stars.

The final spectra are not corrected for foreground dust extinction. Users should make these corrections before using them.

\subsubsection{Quality Control} \label{sec:mjdqual}

A stellar library requires strict quality control. We have a number of quality assessment carried out in the pipeline to flag poor quality spectra. We identify cases having low S/N, bad sky subtraction, high scattered light, low PSF-covering fraction, uncertain radial velocity measurement, and/or those with problematic flux calibration. Each spectrum we release has an associated quality bitmask ({\tt MJDQUAL} for each visit spectrum) giving these quality information. We provide a summary of these in Table \ref{table:mjdqual} and describe them in more detail here. 

 \begin{deluxetable*}{ll}
\tablecaption{Qualitiy Control Bits {\tt MJDQUAL} for MaStar. See \S \ref{sec:mjdqual} for full explanation. \label{table:mjdqual}}
\tablehead{\colhead{Bit} & \colhead{Description} } 
\startdata
1 & Problematic sky subtraction \\
2 & High scattered light in the raw frame \\
4 & Low PSF covering fraction \\ 
5 & Poor flux calibration  \\
6 & Unreliable radial velocity estimates \\
7 & Flagged as unreliable from visual inspection \\
8 & Strong emission lines \\
9 & Low $S/N$
\enddata
\end{deluxetable*}

A large fraction of the observed stars have problematic flux calibration due to the fact that the standard stars on those plates have much less extinction than given by the \citet[][SFD]{SFD} dust map. In the flux calibration step of the MaNGA pipeline, we assume the standard stars are behind the Galactic dust, and we compare the observed spectra of the standard stars with the dust-extincted theoretical models to derive the instrument throughput curve. This assumption is valid for all galaxy plates which are at high Galactic latitude and have relatively faint standard stars, placing them at a safe distance behind most of the dust. However, this assumption fails for many fields targeted by MaStar, which are at low Galactic latitudes. Stars at low Galactic latitude are quite likely to be found in front of some fraction of the dust in that direction. Thus, when applying the extinction given by SFD, we overly redden the theoretical models and arrives at an incorrect flux calibration for these field. We have a solution to this problem which will be incorporated into the MaStar pipeline in the future. In the current release, the spectra for these stars are just flagged as having poor flux calibration (bit 5 of {\tt MJDQUAL}).  

A significant fraction of the spectra also have unreliable radial velocity estimates. We used a rather limited set of templates in our derivation of the radial velocities. Thus stars with very hot or very cool effective temperature, and those with very high and low surface gravity, are more likely to be affected by this issue. These can be identified by checking bit 6 of the {\tt MJDQUAL} bitmask. All spectra are shifted to rest-frame according to the reported heliocentric radial velocity, regardless of whether the measurement is robust or not.

In addition to these automated checks, we carried out a visual inspection campaign to ensure the quality of each spectrum. Using the Zooniverse Project Builder interface\footnote{\url{https://www.zooniverse.org/lab}}, we started a private project for visually inspecting the spectra. A total of 28 volunteers from within the collaboration participated in the campaign, 10,797 visit spectra were inspected, each by at least 3 volunteers, to check for issues in flux calibration, sky subtraction, telluric subtraction, emission lines, etc. The results are input to the data reduction pipeline to assign the final quality flags. 

The primary set of spectra we are releasing contains only those spectra that are deemed to have sufficient quality to be useful. We have excluded from the primary set those spectra with problematic sky subtraction (bit 1 of {\tt MJDQUAL}), low PSF covering fraction (bit 4), poor flux calibration (bit 5), or low S/N (bit 9), and those identified as problematic by visual inspection (bit 7). The primary set still contains spectra with unreliable heliocentric velocity measurement (bit 6), whose spectra would still be in the observed frame. It also contains spectra with strong emission lines (bit 8), some of which are intrinsic to the star. In addition, stars flagged to have high scattered light in the raw frame (bit 2) are also included as they may not be affected significantly. Other bits that are not mentioned above were never set in the current data release. The users are strongly advised to check the quality flags when using the spectra. Detailed information about the quality flags can be found in \citet{Yan2018}.  

In addition to these basic quality checks, we are testing the spectra by running them through a population synthesis code \citep{Maraston2005}. This procedure, which will be described in C. Maraston et al. (in prep)., allows us to test the total effect of goodness of spectra plus assigned stellar parameters and will be crucial for the joint calculation of stellar parameters and stellar population models. Relevant to this description, this method allows us to spot bad or highly-extincted spectra.

\subsubsection{Stellar Parameter Distribution}

Robust assignment of stellar parameters to the stars are also critical for the stellar library. Our targets are selected from heterogeneous sources. Those selected from APOGEE, SEGUE, and LAMOST have parameters available from their respective catalogs. However, they are measured with different methods and may not be consistent with each other. They also have different boundaries applied in the determination of the parameters. For our target selection purpose, we have made small constant corrections to the parameters to remove the overall systematic difference. These slightly adjusted parameters are included in the catalog we release. However, the correction are done independent of detailed stellar types. As a result, the parameters from different catalogs can still have subtle stellar-type-dependent systematic differences. For individual stars, they could be used to determine the rough stellar type. But for the library as a whole, we caution against using these input parameters to compare the stars or to construct stellar population models with them.

We are still in the process of determining stellar parameters for all the stars in the MaStar library, in a way that is as homogeneous as possible. This is not an easy task, because for stars with different stellar types we need to rely on different spectral features and different methodology. Although these are not yet available in this version of the library, we present here the extinction-corrected Hertzsprung-Russell (HR) diagram for our stars using photometry and parallax from Gaia DR2 \citep{GaiaDR2,Evans2018} and Gaia-parallax-based distance estimates from \citet{Bailer-Jones2018}. This provides a rough idea of our stellar parameter coverage. Here we only plot stars that are either in directions with a total $E(B-V)$ less than 0.1 mag or more than 300pc above or below the Milky Way mid-plane so that we can use the total amount of dust measured by \citet{SFD} for the extinction correction reliably. The photometry of our targets also come from various sources including PanSTARRS1 \citep{Chambers2016}, APASS\footnote{\url{https://www.aavso.org/apass}}, SDSS, Gaia DR1 \citep{GaiaDR1}, and Tycho-2 \citep{Hog2000} for a few stars. For stars with PanSTARRS1 photometry, we converted them to SDSS using the formula provided by \citet{Finkbeiner2016}. For APASS, we assume they are in SDSS filters already. For all the other non-SDSS stars, we use Gaia DR2 photometry to derive the magnitudes in SDSS $gri$ bands according to the conversion given by \citet{Evans2018}. In Figure~\ref{fig:mastar}, we show the $r$-band absolute magnitude ($M_r$) vs. $g-i$ for these stars. The color-coding are based on our preliminary measurement of metallicity using the ULySS pipeline \citep{Koleva09, Koleva11} with MILES \citep{MILES} as the training set. 

\begin{figure}[htp]
\begin{center}
\includegraphics[width=0.45\textwidth]{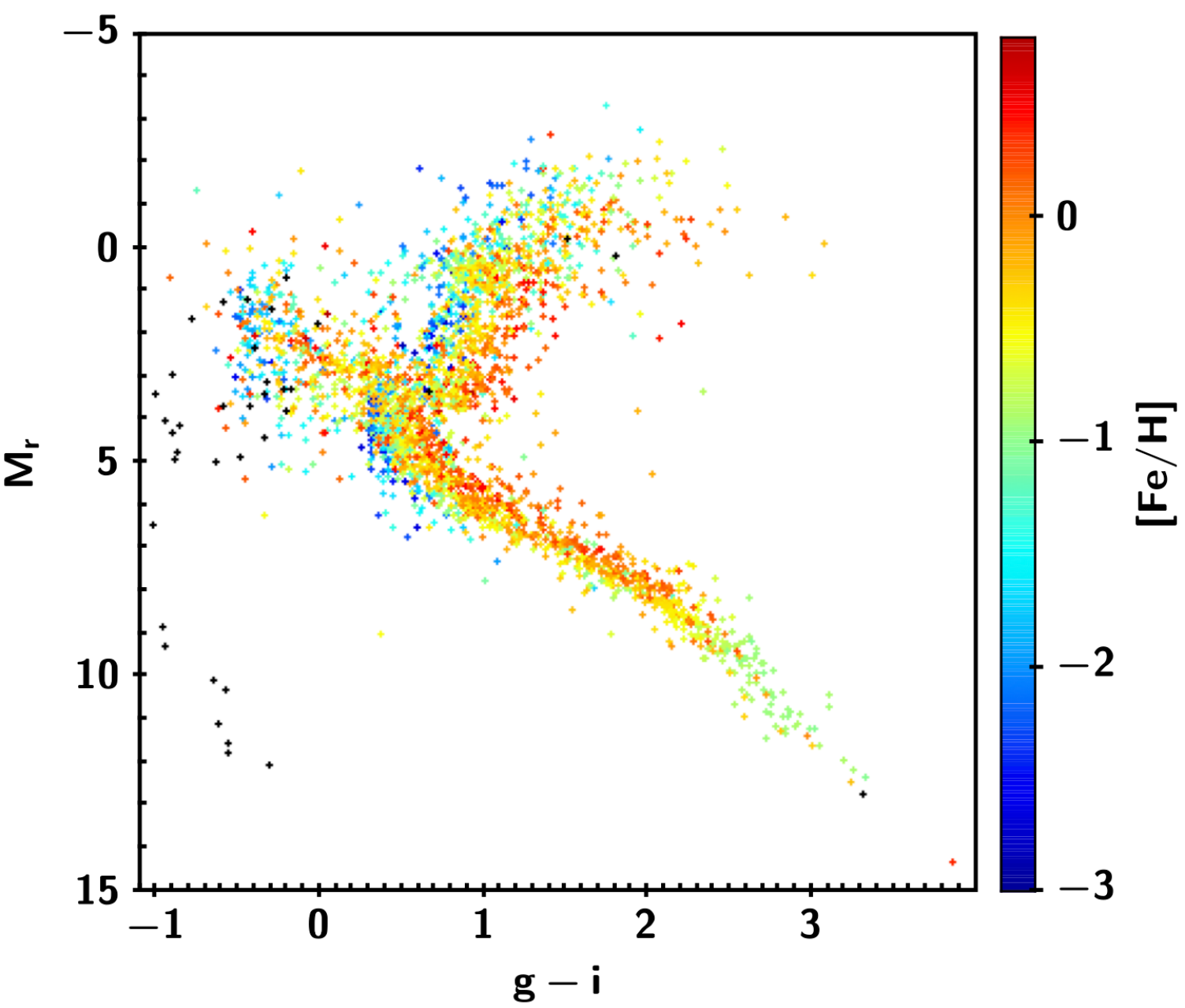}
\end{center}
\caption{Extinction-corrected HR diagram for MaStar targets, color-coded by our preliminary measurement of metallicity. The $g$, $r$,and $i$ bands are in the SDSS photometry system. The absolute magnitudes are derived using parallax-based distance estimates from \citet{Bailer-Jones2018}. Only stars for which we are able to get an approximate extinction correction are included here. This Figure is reproduced from \citet{Yan2018}.}
\label{fig:mastar}
\end{figure}

From Figure~\ref{fig:mastar}, we can see that the current released subset of MaStar library already has a very good coverage across the HR diagram for a wide range of metallicities. But there is significant room for improvement. We need to cover more stars at the luminous end of the red giant branch and at the blue end of the main sequence. These will be added in future versions of the library.

\subsubsection{Data Access and usage information}

The MaStar data for this release can be found in two main files: the ``{\tt mastarall}'' and the ``{\tt mastar-goodspec}'' files. Both files can be found on the SAS (\url{https://dr15.sdss.org/sas/dr15/manga/spectro/mastar/v2_4_3/v1_0_2/}, with datamodels at \url{https://data.sdss.org/datamodel/files/MANGA_SPECTRO_MASTAR/DRPVER/MPROCVER/}. 

The ``{\tt mastarall}'' file contains four summary tables of the basic information about the stars and of the information about their one or more visits. They include the identifier (MaNGAID), astrometry, photometry, targeting bitmask which indicates source of photometry, input stellar parameters, plate, IFU, modified Julian date of the visit, derived heliocentric velocity, and spectral quality information. MaNGAID is the identifier to identify unique stars, except in a few cases where the same star was assigned two different MaNGAIDs. These are documented in detail online and in \citet{Yan2018}. 

The ``{\tt mastar-goodspec}'' file contains the primary set of high quality visit spectra. In addition to identification information, we provide the wavelength, flux, inverse variance, mask, spectral resolution vector, and the spectral quality bitmask. 

MaStar spectra can also be visualized in the SAW (see \S \ref{sec:dataaccess} for details). 

\subsection{Other Ancillary Programs}
\label{sec:mangaancillary}

 We refer the reader to the DR13 paper \citep{2017ApJS..233...25A} for the most complete list of MaNGA ancillary programs\footnote{also see \url{http://www.sdss.org/dr15/manga/manga-target-selection/ancillary-targets}}. These approved programmes make use of $\sim$5\% of MaNGA bundles, and provide in Table \ref{table:ancillary} an updated list of the number of bundles available in each documented sample. 
 
 \begin{deluxetable*}{lcrc}
\tablecaption{Summary of MaNGA Ancillary Programs with Data in DR15\label{table:ancillary}}
\tablehead{\colhead{Ancillary Program} & \colhead{Observed\footnote{These are bundle counts, not always unique galaxies}} & \colhead{{\tt BITNAME}} & \colhead{binary digit}} 
\startdata
Luminous AGN & 25\footnote{Count for 1,2,3,4 combined} & {\tt AGN\_BAT} & 1 \\
 & & {\tt AGN\_OIII} & 2 \\
 & & {\tt AGN\_WISE} & 3 \\
 & & {\tt AGN\_PALOMAR} & 4 \\
Void Galaxies & 3 & {\tt VOID} & 5\\
Edge-On Star-forming Galaxies & 20 & {\tt EDGE\_ON\_WINDS} & 6 \\
Close Pairs and Mergers & 57\footnote{Count for 7,8,9,10 combined} & {\tt PAIR\_ENLARGE} & 7 \\
& & {\tt PAIR\_RECENTER} & 8 \\
& & {\tt PAIR\_SIM} & 9 \\
& & {\tt PAIR\_2IFU} & 10 \\
Writing MaNGA (public outreach) & 1 & {\tt LETTERS} & 11\\
Massive Nearby Galaxies  & 23 & {\tt MASSIVE} & 12 \\
Milky Way Analogs & 4 & {\tt MWA} & 13 \\
& 0 & {\tt MW\_ANALOG} & 23 \\
Dwarf Galaxies in MaNGA & 22 & {\tt DWARF} & 14 \\
Brightest Cluster Galaxies & 24 & {\tt BCG} & 17 \\
MaNGA Resolved Stellar Populations & 1 & {\tt ANGST} & 18 \\
Coma & 68 & {\tt DEEP\_COMA} & 19 \\
IC 342 & 50 & {\tt IC342} & 20 \\
M31 & 18 & {\tt M31} & 21 \\
SNIa Hosts & 1 & {\tt SN1A\_HOST} & 26
\enddata
\end{deluxetable*}

 There are three new ancillary programs for DR15. These provide observations in fields of IC 342, M31 as well as data for a selection of SN1a Hosts ({\tt MANGA TARGET3} target bits 20, 21 and 26, respectively). 
 
 The IC342 program will uniformly mosaic the disk of IC 342 using 61 plates, following an initial pilot program of 3 plates that target individual HII regions across the disk.  This galaxy serves as a local reference with 30 pc resolution that can inform our understanding of the unresolved physics in the $\sim1$ kpc resolution main MaNGA survey.
 
 The M31 ancillary program targets regions in M31 where the underlying physical properties are well-constrained from resolved stellar population analyses, provided by the Panchromatic Hubble Andromeda Treasury (PHAT; \citealt{dalcanton2012}).  The MaNGA observations include 18 regions (~50-100 pc in size) that sample a wide range of environmental conditions, including ancient and recent star formation history, dust column, dust geometry, and metallicity. These observations provide a link between resolved stellar populations and the inferred properties of unresolved stellar populations, and can be used to assess the ability of spectral fitting codes to recover key physical parameters.
 
 The SNIa Hosts ancillary program will observe Type Ia supernova (SNIa) host galaxies to investigate causes of the intrinsic variation of SNIa. SNIa show a spread in absolute magnitude, but can be standardized by taking into account relationships like luminosity-decline rate and SNIa color to reduce the spread to 0.12 mag. Research over the past several years indicates that some of this remaining spread correlates with global host galaxy properties such as stellar mass, star-formation history, and metallicity (e.g. \citealt{Lampeitl2010, Gupta2011, Hayden2013, Rigault2013}) causing concerns about biases in cosmological measurements. This project will obtain MaNGA data for roughly 40 SNIa host galaxies in order to look for correlations with SNIa peak absolute magnitude and host galaxy properties like metallicity and star formation rate averaged over the whole galaxy and at the location of the SNIa.

 DR15 also incorporates a second Milky Way Analogs program (target bit 23), which is similar to the existing program described in DR13, but uses morphological information, rather than star formation rates, in combination with galactic stellar mass to select analogs.
 
\subsection{Value Added Catalogs}

As was the case previously in DR14, there are a large number of Value Added Catalogs (VACs) linked to MaNGA data released in DR15. These either represent additional processing of DRP or DAP output, or follow-up programs or other data useful in combination with MaNGA data. We summarize new or updated VACs below. 

\subsubsection{Spectral Modeling \label{mangaspec}}

In DR15 there are new releases for both the {\tt FIREFLY} \citep{goddard2017} and {\tt Pipe3D} \citep{sanchez2016b} stellar population modeling and emission line analysis VACs. Full details of both can be found in the DR14 paper (and references therein), so we give only updates specific to this DR15 release version below.

 For DR15 {\tt Pipe3D} version 2.4.3 was run over the MaNGA DR15 dataset. The main difference with respect to version 2.1.2 (used in DR14) besides the number of analyzed galaxies, was to solve a bug in derivation of the equivalent widths of the analyzed emission lines. The current {\tt Pipe3D} VAC provides two different types of data products: (1)  A catalog comprising 94 different parameters measured for each of 4660 galaxies (all galaxies in MaNGA cubes for which {\tt Pipe3D} was able to derive the main stellar population, emission line and kinematics properties); and (2) a set of 4660 datacubes {\tt manga.Pipe3D.cube.fits} presenting a set of spatially resolved parameters. The parameters are the same as they were in the DR14 version \citep{sanchez2018}. More detail is available on the data release website \url{https://www.sdss.org/dr15/manga/manga-data/manga-pipe3d-value-added-catalog/}.

The major update to {\tt FIREFLY} with respect to DR14 is the extension of the stellar population modeling grid based on the models of \citet{Maraston2011}. The new catalog uses a finer metallicity grid with the following grid values: $[Z/{\rm H}] = -2.3, -1.9, -1.6, -1.2, -0.9, -0.6, -0.3, 0.0, 0.3$. The new version of the VAC also provides geometrical information so that maps can be produced directly from the VAC (a python plotting routine is available from the data release website). The entire VAC is available as either a single fits file containing all measurements, or smaller fits files with selected subsets of the derived parameters. More detail on the catalog is provided on the data release website \url{https://www.sdss.org/dr15/manga/manga-data/manga-firefly-value-added-catalog/} and in \citet{goddard2017} and \citet{parikh2018}.

\subsubsection{Morphology and Photometry of MaNGA Targets}
\label{Mangamorph}

As part of DR15 we release one photometry VAC and two morphology VACs. 

The {\tt PyMorph} catalog provides photometric parameters obtained from Sersic and Sersic+Exponential fits to the 2D surface brightness profiles of the MaNGA DR15 galaxy sample. It uses the {\tt PyMorph} algorithm for determining the fits which has been extensively tested \citep{meert2013,fischer2017,bernardi2017}, and {\tt PyMorph} reductions of SDSS DR7 galaxies \citep{DR7} are available (the UPenn SDSS PhotDec Catalog: \citealt{meert2015, meert2016}). We have re-run {\tt PyMorph} for all the galaxies in the MaNGA DR15 sample. These re-runs incorporate three improvements: they use the SDSS DR14 images, improved bulge-to-disk decomposition by slightly modifying our criteria when using {\tt PyMorph} (see \citealt{Fischer2018}. for details), and all the fits in this catalog have been visually inspected for additional reliability (we recommend using ``{\tt flag\_fit}''). The catalog contains these fits for the g, r, and i bands. One important caveat to note is that position angles (PA) are reported relative to the SDSS imaging camera columns, which are not aligned with North, so a correction is needed to convert to true position angles. To convert to the usual convention where North is up, East is left (note that the MaNGA datacubes have North up, East right) set ${\rm PA}_{\rm MaNGA} = 90\deg-{\rm PA}_{\rm PyMorph} - {\rm PA}_{\rm SDSS}$, where ${\rm PA}_{\rm PyMorph}$ is the value given in this catalogue, and ${\rm PA}_{\rm SDSS}$  is the SDSS camera column position angle with respect to North.

 A curated version of the Galaxy Zoo crowdsourced classifications containing an entry for all MaNGA target galaxies is released in DR15. This catalog contains galaxy classifications previously released in \citet[which was selected from the SDSS DR7 galaxy catalog]{GZ2} as well as new unpublished classifications for MaNGA targets missing from that list. All morphological identifications are provided based on the citizen scientist input using the improved technique for aggregation and debiasing described in \citep{Hart16}. This accounts better for redshift bias in the detailed classifications of spiral arms, bars than the version used in \citet{GZ2}. For a simple conversion between Galaxy Zoo classifications and modern (bulge-sized based) T-types see details in \citet{GZ2}, which also includes general advice on how to best use Galaxy Zoo classifications for science. 

A second morphology catalog is provided that has been obtained with the help of ``Deep Learning" models. The models were trained (making use of Galaxy Zoo morphologies, as well as morphologies from \citealt{nair2010a}) and tested on SDSS-DR7 images \citep{dominguezsanchez2018}. The morphological catalog contains a series of Galaxy Zoo like attributes (edge-on, barred, bulge prominence and roundness), as well as a T-Type and a finer separation between pure elliptical and S0 galaxies.

\subsubsection{HI-MaNGA - HI 21cm Follow-up for MaNGA}
\label{himanga}

The first data release of ``HI-MaNGA", the HI followup project for MaNGA is provided as a VAC in DR15. This follow-up program is presented in \citet{Masters2018} and is the result of single dish radio 21cm HI observations of MaNGA galaxies using the Robert C. Byrd Green Bank Telescope (GBT). The depth of this observing is aimed to be similar to the Arecibo Legacy Fast Arecibo L-band Feed Array (ALFALFA) blind HI survey \citep{haynes2018} which covers some of the MaNGA footprint (see Figure \ref{fig:mangaforecast}) with a goal of enabling studies to use HI data from both surveys. In this first release, data are provided for 331 MaNGA galaxies observed in the 2016 GBT observing seasons under project code {\tt AGBT16A\_95}. Total HI masses, and line widths (measured with five different common techniques) are provided for all detections, while HI mass upper limits (assuming a line width of 200km/s) are provided for non-detections. HI-MaNGA has observed an additional $\sim 1800$ MaNGA galaxies in the 2017 observing season (under project code {\tt AGBT17A\_12}); these data will be released in a future VAC. The sky distribution of all MaNGA galaxies observed by this program is shown in Figure \ref{fig:mangaforecast}. 

\subsubsection{GEMA-VAC; Galaxy Environment for MaNGA Value Added Catalog} 
\label{manga:gema}

The environment in which a galaxy resides plays an important role in its formation and evolution. Galaxies evolve as a result of intrinsic processes (i.e. their nature - this includes processes such as internal secular evolution, feedback of various kinds etc), but they are also exposed to the influences of their local and large-scale environments (i.e. how they are ``nurtured''). We present a the Galaxy Environment for MaNGA Value-Added Catalog (GEMA-VAC) which provides a quantification of the the local and large-scale environments of all MaNGA galaxies in DR15. There are many different definitions of environment and there are also several ongoing projects within the MaNGA team exploring the influence the environment on galaxy properties. With this VAC we aim to join and coordinate efforts so that the entire astronomical community can benefit from the products. The GEMA-VAC catalog will be described in more detail in M. Argudo-Fern\'andez et al (in prep). We describe the contents of the VAC briefly below. 

We estimate the tidal strength parameter for MaNGA galaxies in pairs/mergers (B. Hsieh et al., in prep.), the tidal strengths exerted by galaxies in the catalog of galaxy groups in \citet{2007ApJ...671..153Y}, and tidal forces exerted by nearby galaxies in two different fixed aperture volume limited samples (namely 1 and 5 Mpc projected distances within a line-of-sight velocity difference of $\Delta\,v~\leq~500$\,km\,s$^{-1}$ \citealt{2015A&A...578A.110A}). Estimations of the local densities with the distance to the 5$^{th}$ nearest neighbor are also provided. The local density within N nearest neighbors include the corrections explained in \citet{2017MNRAS.465..688G} following the methodology described in \citet{2015MNRAS.451..660E}. To have a more general picture of the environment, we also provide a characterization of the cosmic web environment which can be used to identify galaxies in clusters, filaments, sheets, or voids, as explained in \citet{2017MNRAS.465.4572Z}. The full details of the reconstruction of these density and tidal fields are described in \citet{2009MNRAS.394..398W} and \citet{2012MNRAS.420.1809W}. 

\subsubsection{MaNGA Spectroscopic Redshifts}
\label{manga:specz}

We present a Value-Added Catalog that contains the best-fit spectroscopic redshift and corresponding model flux for each MaNGA spectrum that has sufficient signal-to-noise. We provide the mean of the spectroscopic redshifts sampled within the inner high signal-to-noise region of the MaNGA galaxies, which can be compared to the single-valued NSA catalog redshift.  Since the MaNGA instrument uses the BOSS spectrograph, our results are derived by iterative application of the BOSS pipeline's spec1d software \citep{2012AJ....144..144B} to precisely measure the redshift using the higher signal-to-noise measurements as a prior to the algorithm, which allows us determine good redshifts on spectra that would not otherwise have sufficient signal-to-noise to result in a good redshift. Since the MaNGA survey uses an IFU, the radial velocity profile of the galaxy can significantly impact the redshift of each spectra. Thus the spectroscopic redshifts can both be a benefit to galaxy kinematic measurements and improve the accuracy of spectra modeling and analysis.  The authors of this Value-Added Catalog have used the spectroscopic redshifts to search for background emission lines to discover strong gravitational lenses in MaNGA \citep{2018MNRAS.477..195T}.

\section{Other Survey Data and Products} 
\subsection{APOGEE-2}
\label{sec:apogee}
SDSS DR15 includes {\it no new} APOGEE data.  The currently available set of APOGEE Survey data consists of the first two years of SDSS-IV APOGEE-2 (Jul 2014-Jul 2016) as well
as the entirety of SDSS-III APOGEE-1 (Aug 2011-Jul 2014) and is an exact {\it duplicate} of that data which was released in DR14.  DR15-associated APOGEE documentation builds upon that from DR14 with extended explanations and the addition of information and relevant text (e.g., a description of the New Mexico State University (NMSU) 1.0m Telescope).  Note that the DR15 APOGEE data model has remained largely the same with only slight revisions to the text for clarity.  Described below are the APOGEE technical papers that contain details which should assist users in the exploitation of APOGEE data as well as provide further understanding as to data quality \citep{Zasowski_2017_apogee2targeting,Holtzman2018,jonsson2018,Pinsonneault2018,Wilson2018}.  Additionally, details are provided on the recently-generated Value-Added Catalog (VAC) from \citet{Donor2018}, which contains a catalog of identified APOGEE open cluster members. There are currently 4 VACs that rely upon APOGEE DR14 in order to extend and enhance the standard APOGEE data release products (DR14 APOGEE TGAS Catalog, APOGEE Red Clump Catalog, APOGEE DR-14 Based Distance Estimations, and OCCAM).\footnote{More information regarding all available APOGEE VACs including brief descriptions and the corresponding authors may be found in the SDSS online documentation (\url{http://www.sdss.org/dr15/data_access/value-added-catalogs/}).}   

\subsubsection{Technical Papers}
Two new APOGEE-related technical papers are highlighted below: the instrument paper from \citet{Wilson2018} that relays an extensive description of the APOGEE spectrographs, and the APOKASC paper from \citet{Pinsonneault2018} that details the APOGEE spectroscopic follow-up of Kepler stars. 

\vspace{0.1in}
{\bf APOGEE Instrument Paper}\\
The forthcoming publication from \citet{Wilson2018} describes the design and performance of the near infrared, fiber-fed, multi-object, high resolution APOGEE spectrographs.  Since 2011, the first APOGEE instrument has been in operation on the 2.5-m Sloan Telescope at the Apache Point Observatory in New Mexico, USA (a Northern hemisphere site).  Several key innovations were made during the development of the APOGEE instrument which include a multi-fiber connection system known as a 'gang-connector' which allows for the simultaneous disconnection and reconnection of 300 fibers; hermetically sealed feedthroughs to permit fibers to pass through the cryostat wall continuously; the first cryogenically-deployed mosaic volume phase holographic grating; and, a massive refractive camera that is comprised of large-diameter mono-crystalline silicon and fused silica elements.  Specifically for the Northern spectrograph, \citet{Wilson2018} reports on the following: the performance of the 2.5-m Sloan Foundation Telescope in the near infrared wavelength regime; the cartridge and fiber systems; the optical and optomechanical systems; the detector arrays and electronic controls; the cryostat; the instrument control system; calibration procedures; instrument optical performance and stability; and, lessons learned.  The final sections of \citet{Wilson2018} provide similar details on the second APOGEE spectrograph located at the 2.5-m du Pont Telescope at Las Campanas Observatory in Chile.  This second (Southern hemisphere-based) instrument, a close copy of the first, has been operating since April 2017.  Wilson et al. also contains multiple appendices for the interested user. 

\vspace{0.1in}
{\bf The Second APOKASC Catalog}\\
Over both the APOGEE-1 and APOGEE-2 Surveys, a joint effort known as the APOGEE Kepler Asteroseismic Science Consortium (APOKASC) APOGEE has engaged in a spectroscopic follow-up of stars in the Kepler field.  Pinsonneault et al (submitted) presents the second APOKASC Catalog of stellar properties for a sample of 6681 evolved stars with APOGEE spectroscopic parameters and \textit{Kepler} asteroseismic data analyzed using five independent techniques.  The APOKASC data includes evolutionary state, surface gravity, mean density, mass, radius, age and the spectroscopic and asteroseismic measurements used to derive them.  As shown in Figure \ref{fig:apogee_apokasc}, the APOKASC catalog asteroseismic log g values and evolutionary state classifications allow for a clear distinction between Red Giant Branch (RGB) and Red Clump (RC) members.  \citet{Pinsonneault2018} employ a new empirical approach for combining asteroseismic measurements from different methods, calibrating the inferred stellar parameters, and estimating uncertainties. With high statistical significance, they find that asteroseismic parameters inferred from the different pipelines have systematic offsets that are not removed by accounting for differences in their solar reference values. Pinsonneault el al. includes theoretically motivated corrections to the large frequency spacing ($\Delta \nu$) scaling relation as well as calibrates the zero point of the frequency of maximum power ($\nu_{\rm max}$) relation to be consistent with masses and radii for members of star clusters. For most targets, the parameters returned by different pipelines are in much better agreement than would be expected from the pipeline-predicted random errors, but 22\% of them had at least one method not return a result and a much larger measurement dispersion. This supports the usage of multiple analysis techniques for asteroseismic stellar population studies.

\begin{figure}
\centering
\includegraphics[angle=0,width=8.8cm]{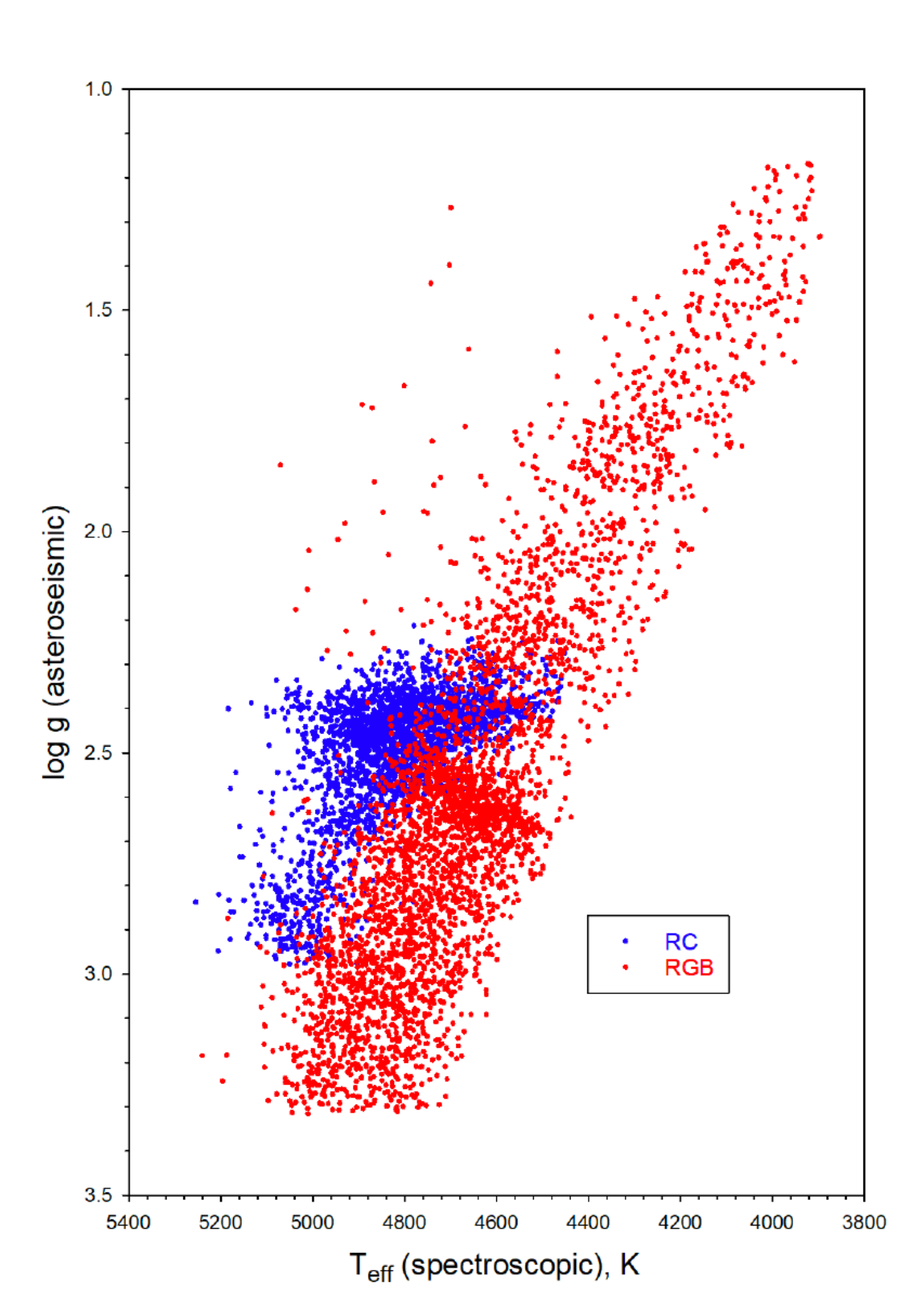}
\caption{Spectroscopic effective temperature (from APOGEE DR14) versus asteroseismic surface gravity (log g) in the APOKASC sample by asteroseismic evolutionary state. Red clump (RC; core He-burning) stars are signified in blue while Red Giant Branch (RGB; H-shell or double shell burning) stars are shown in red. The two populations have a clear offset in this plot, with RC stars having higher surface gravity at the same temperature compared to RGB stars.}
\label{fig:apogee_apokasc}
\end{figure}

\vspace{0.1in}
In the SDSS DR14 data release paper \citep{2018ApJS..235...42A}, brief references were made to the \citet{Holtzman2018} and \citet{jonsson2018} publications.  For the benefit of users, concise descriptions of each are now provided.  Please note that in addition to the DR15 Documentation, users should refer to these publications for detailed information regarding the Data Reduction Pipeline (DRP) and the APOGEE Stellar Parameter and Chemical Abundance Pipeline (ASPCAP) as well as to understand data quality and performance.

\vspace{0.1in}
{\bf SDSS/APOGEE DR13 and DR14 Pipeline Processing and Data Description}\\
\citet{Holtzman2018} describes the data and analysis methodology used for the SDSS/APOGEE Data Releases 13 and 14 as well as highlights differences from the DR12 analysis presented in \citet{Holtzman2015}. For example, the work demonstrates some improvement in the handling of telluric absorption and persistence in the DR13/DR14 versions of APOGEE-2 data as opposed to DR12. \citet{Holtzman2018} details the derivation and calibration of stellar parameters, chemical abundances, and respective uncertainties, along with the ranges over which calibration was performed. The work reports some known issues with the public data related to the calibration of the effective temperatures (DR13), surface gravity (DR13 and DR14), and C and N abundances for dwarfs (DR13 and DR14).  \citet{Holtzman2018} also discusses how results from The Cannon \citep{Cannon} are included in DR14 and compares those with the values from ASPCAP. 

\vspace{0.1in}
{\bf Comparison of SDSS/APOGEE DR13 and DR14 Values to Optical Results}\\
\citet{jonsson2018} evaluates the ASPCAP performance for both the DR13 and DR14 APOGEE datasets with 160,000 and 270,000 stars, respectively.  A comparison of the ASPCAP-derived stellar parameters and abundances is done to analogous values inferred from optical spectra and analysis with a subset of several hundred stars.  For most elements, \citet{jonsson2018} find that the DR14 ASPCAP  results have systematic differences with the comparison samples of less than 0.05 dex (median) and random differences of less than 0.15 dex (standard deviation).  These departures are attributed to a combination of the uncertainties in both the comparison samples as well as the ASPCAP-analysis.  Specifically, in comparison to the optical data, \citet{jonsson2018} find that magnesium is the most accurate alpha-element derived by ASPCAP while nickel is the most accurate Fe-peak element (excluding iron).

Additionally in \citet{2018ApJS..235...42A}, detailed information was provided regarding the recently-published APOGEE-2 Targeting Paper from \citet{Zasowski_2017_apogee2targeting}   Users are encouraged to consult Zasowski et al. for specific details and insight regarding APOGEE-2 targeting.

\subsubsection{New Value-Added Catalog - OCCAM \label{apogeevac}}
The Open Cluster Chemical Analysis and Mapping (OCCAM) Survey generates a VAC of open cluster members as targeted in both APOGEE-1 and APOGEE-2 fields.  To establish membership probabilities, the catalog combines APOGEE DR14 DRP-derived radial velocities (RV) and ASPCAP-derived metallicities with proper motion (PM) data from Gaia DR2. This first VAC from the OCCAM Survey includes 19 open clusters, each with 4 or more APOGEE members.  The OCCAM VAC consists of two components: a set of bulk cluster properties which include motions (RV, PM) as well as robust average element abundance ratios; and, a set of membership probabilities for all stars considered in the analysis of the 19 open clusters.  For further information on the OCCAM Survey please consult \citet{Donor2018}.

\subsection{eBOSS, TDSS and SPIDERS}
\label{sec:eboss}

There are no new reduced eBOSS data included in
this data release; the VACs which are released are based on previously released eBOSS spectra.
The final eBOSS spectroscopic sample will be released in DR16. For more details on what's coming in DR16 see \S \ref{sec:ebossfuture}. 

DR14 marked the first cosmological sample from eBOSS, consisting of spectra
predominantly of luminous red galaxies (LRG) and quasars.
These data enabled the first baryon acoustic oscillation (BAO) measurement in the
$1<z<2$ redshift range from quasars \citep{ata18a}
and a 2.6\% precision constraint on the distance scale using the
clustering of LRG's \citep{bautista17b}.
These measurements reflect the two primary goals for early eBOSS
science, yet are only a subset
of the results from the two-year eBOSS sample.  The large-scale
structure catalogs for both of these studies were released in July 2018
and were not described in the 14th Data Release publication.   These
value-added catalogs can now be accessed from the DR14
site\footnote{\url{https://data.sdss.org/sas/ebosswork/eboss/lss/catalogs/DR14/}}
and from this new release in the parallel location.  These catalogs
contain all necessary information such as the window function,
systematic quantities, completeness estimates and corrections for
close-pairs and redshift failures to reproduce those clustering
measurements, similar to the catalogs from the final BOSS sample
\citep{reid16a}.

The publications that document the DR14 target selection algorithms
\citep{myers15a,prakash16a,palanque-delabrouille16a} will also describe
the LRG and quasar samples for the final eBOSS sample.
Several new algorithms for the spectroscopic data reductions were
implemented in DR14 \citep{hutchinson16a,jensen16a}; we will further improve sky subtraction with higher order models to the
fiber-to-fiber sky model, flux calibration with new models for standard
stars, and spectral extraction to account for cross-talk such as that
found in \citep[submitted]{Hemler2018} for the final sample.  A new method to improve the classification of
galaxy spectra \citep{hutchinson16a}
was implemented in DR14 and new methods for classifying emission line
galaxies (ELG) and quasars are being considered for the final sample.

\subsection{Optical Emission Line Properties and Black Hole Mass Estimates for SPIDERS DR14 Quasars}
\label{SPIDERS_VAC_lines}
This VAC, released in DR15, contains optical spectral properties 
for all X-ray selected SPIDERS quasars released in DR14. 
The SPIDERS DR14 catalog is based on a clean sample 
of 9399 sources from the Second ROSAT All-Sky Survey catalog
\citep[2RXS;][]{2016A&A...588A.103B}
and 1413 sources from the first XMM-Newton Slew survey catalog 
\citep[XMMSL1;][]{2008A&A...480..611S}
with optical spectra available. 
X-ray sources were matched to ALLWISE infrared counterparts 
using the Bayesian algorithm ``NWAY'' \citep{2018MNRAS.473.4937S}, 
which were then spectroscopically identified using SDSS \citep{2017MNRAS.469.1065D}. 
Visual inspection results for each object in this sample are available from 
a combination of literature sources and the SPIDERS group, 
which provide both reliable redshifts and source classifications. 
A spectral fitting code has been produced which fits the spectral regions 
around the H$\beta$ and MgII emission lines
and provides both line and continuum properties, 
bolometric luminosity estimates, 
as well as single-epoch black hole mass estimates.
This VAC includes X-ray flux measurements, visual inspection results, 
optical spectral properties, black hole mass estimates, 
and additional derived quantities for all SPIDERS DR14 quasars. For more details see  \citet{Coffey2018}. 

\section{Future Plans}
\label{sec:future}

SDSS-IV has a full 2 years of operations remaining, and is planning a further two public data releases. The next data release, DR16, is now scheduled for December 2019 and will comprise data taken by both the APOGEE-N and APOGEE-S instruments through July 2018 as well as being the final complete data release for eBOSS operations. The final, complete release, DR18 (which will follow an internal only DR17) is planned for December 2020. 

\subsection{eBOSS} \label{sec:ebossfuture}
The eBOSS schedule was recently accelerated in order to achieve
its cosmological goals earlier than previously planned. This
acceleration began on January 1, 2018 and continues through February
16, 2019, at which time eBOSS will complete its program significantly
ahead of the start of the Dark Energy Spectroscopic Instrument (DESI) survey \citep{DESI1}. Under the original schedule, eBOSS and MaNGA
divided the dark time roughly equally. Under the new schedule, eBOSS
will control all the dark time in January and February 2018, will
return to the original schedule for March through July 2018, and will
control all the dark time from August 2018 through February 16, 2019.

The total time scheduled for eBOSS in this period amounts to 1540 hours.
Historical weather patterns indicate that 50\% of this time
will allow for open dome with seeing and transparency conducive to
spectroscopy.  Based on the 1158 plates completed from July 1, 2014 until June 30,
2018, we project
the final data sample to include the spectra from observations
covering 302 ELG plates and more than 1200 LRG/quasar plates.  These
observations will define the complete ELG sample following the selection
algorithms in \citet{raichoor17a}.  The final LRG and quasar
samples will cover a volume roughly 2.3 times larger than the two-year
cosmology samples released and analyzed in DR14.

The final eBOSS sample will enable precision measurements of
BAO in the clustering of galaxies, quasars, and the Lyman-$\alpha$ forest.
The final sample will also enable new measurements of redshift space
distortions
in the anisotropic clustering of galaxies and quasars over the redshift
range $0.6 < z < 2.2$.
The next data release has been scheduled around the expected time that
these analyses will complete.  This data release will be the last to include new eBOSS
data.  Also included will be the
value-added catalogs that allow others to reproduce the final cosmology
measurements.

\subsection{SPIDERS}
At the completion of the eBOSS survey, SPIDERS will have only obtained spectra from the ongoing followup program of ROSAT and XMM-Newton sources. Continuing at the current pace, at the end of the survey SPIDERS will have collected about 12000 new spectra of X-ray selected AGN and 40000 spectra of member galaxies of about 5000 clusters over the final eBOSS area. 

The delayed launch of the {\it eROSITA} satellite \citep{Predehl_2014_eROSITA}, combined with the accelerated program for obtaining eBOSS spectra mean that it will not be possible to obtain redshifts for {\it eROSITA} targets during routine eBOSS operations. The {\it eROSITA} Performance Verification
data set is currently planned to be available by early-mid 2019 and should
consist of 120 sq deg, with 100-140 targets per sq deg. To address at
least part of the original goals of SPIDERS involving {\it eROSITA} followup we plan
to dedicate a special set of 12 plates for these targets, however this plan cannot be confirmed until February 2019. 

\subsection{TDSS}
The accelerated pace for eBOSS discussed above correspondingly accelerates TDSS, which also relies on the BOSS spectrographs, using a
small portion (about 5\%) of the optical fibers piggybacking on eBOSS plates. TDSS observations will thus effectively also conclude with eBOSS
data collection in about mid-February of 2019, and with SDSS-IV/TDSS data to be included in the future DR16. Although all three main components of TDSS -- the optical spectroscopic follow-up of PS1 photometric imaging variables (e.g., see
\citealt{Morganson2015}, \citealt{Ruan2016}), repeat Few-Epoch Spectroscopy
(FES) of selected subclasses of stars and quasars anticipated or suspected
to reveal spectroscopic variability (e.g., see \citealt{MacLeod2018}), and
the more recently initiated TDSS Repeat Quasar Spectroscopy (RQS; also see
\citealt{MacLeod2018}) program -- thereby also have been accelerated toward
completion, in practice this advance is such that SDSS-IV data collection for the TDSS RQS program in particular is now nearing completion\footnote{This is primarily just because RQS piggybacks on a subset of eBOSS plates which received recent heavy emphasis}. The TDSS
RQS program obtains multi-epoch spectra for thousands of known quasars
(and with larger sample size, and greater homogeneity and less a priori
bias to specific quasar subclasses than the TDSS FES programs), all of
which have at least one earlier epoch of SDSS spectroscopy already available
in the SDSS archive. The RQS program especially addresses quasar spectral
variability on multi-year timescales, and in addition to its own potential
for new discoveries of phenomena such as changing-look quasars or broad absorption line (BAL)
variability and others, will also provide a valuable (and timely) resource for
planning of yet larger scale multi-epoch quasar repeat spectral
observations anticipated for the Black Hole Mapper program in the future
SDSS-V (see \S \ref{sdss5} below). From data taken for the RQS SDSS-IV program to date,
we expect RQS to add another recent epoch of spectroscopy for $\sim$16000
SDSS quasars, sampling across a broad range of properties including redshift, luminosity, and
quasar subclass type. 

\subsection{MaNGA}

MaNGA will continue to take observations for the next two years of SDSS-IV operations. The time trade with eBOSS has slowed the rate of observations during 2018; however, it will provide an overall increase in the total observing time allocation for MaNGA by 8\%.  The projected final survey footprint, assuming we continue nominal survey operations through July 2020, is shown for two different expectations for weather at the telescope and overlaid on other relevant surveys in Figure \ref{fig:mangaforecast}. We expect to exceed our original goal of 10,000 galaxies slightly under nominal weather conditions.

\subsection{APOGEE-2}

The APOGEE-2 Survey continues to acquire observations from both the Northern and Southern hemispheres. SDSS-IV Data Release 16 will contain the first APOGEE-2 data from the Southern instrument.  For DR16, a variety of improvements are planned to both the DRP and the APOGEE Stellar Parameters and Chemical Abundance Pipeline (ASCAP).  A new atomic line list will be generated (which will include transitions of Ce II and Nd II) and a new molecular list will be assembled(which will be more extensive in size and will incorporate FeH features).  An expansion of the stellar atmosphere model grid is underway which will entail the inclusion of higher surface gravities, lower carbon abundances, and higher nitrogen abundances.  Note that Model Atmospheres in Radiative and Convective Scheme (MARCS\footnote{http://marcs.astro.uu.se}) models \citep{Gustafsson2008} will be employed for both the M and GK grids, ensuring a smooth transition across an effective temperature range of approximately $T_{eff} = 2500 - 6000 K$.  Additionally, some tweaking of the data processing and derivation procedure will occur.  Planned modifications include the improvement of the LSF determination (with the potential employment of on-the-fly LSF derivation), a better methodology for the extraction of the individual element abundances, and an improved technique for filling holes in the stellar atmosphere model grid.

\subsection{SDSS-V} \label{sdss5}

Preparations for the fifth generation of SDSS are underway, with SDSS-V anticipated to begin operations in 2020  \citep{2017arXiv171103234K} .  SDSS-V will collect data at both APO and LCO using the existing APOGEE and BOSS spectrographs on the 2.5-meter telescopes, as well as new optical spectrographs dedicated to integral field spectroscopy on new smaller telescopes.  The current SDSS plugplate system will be replaced with robotic fiber positioners in the focal planes of the 2.5-meter telescopes.

SDSS-V comprises three primary projects: the Milky Way Mapper, the Black Hole Mapper, and the Local Volume Mapper.  The Milky Way Mapper will use the APOGEE and BOSS spectrographs to observe 4-5 million stars in the Milky Way and Local Group, probing questions of galaxy formation and evolution, stellar astrophysics, and stellar system architecture.  The Black Hole Mapper will use the BOSS spectrographs to measure masses for $\sim$1200 supermassive black holes via reverberation mapping \citep[e.g.][]{Grier2017}, determine spectral variability for $\sim$25,000 quasars, and provide identifications and redshifts for $\sim$400,000 X-ray sources detected by {\it eROSITA} \citep{Predehl_2014_eROSITA}.  The Local Volume Mapper will collect integral field spectroscopy using new, $R \sim 4000$ optical spectrographs coupled to small telescopes at APO and LCO.  These spectra will span $\sim$3000~deg$^2$ of sky in the Milky Way midplane, the Magellanic Clouds, and other Local Group galaxies at high spatial resolution, with the goal of tracing ISM physics and stellar-ISM energy exchange on different physical scales in a range of galactic environment.

\subsection{Long Term Sustainability of the SDSS Archives}

Starting in 2017, the Science and Catalog Archive Teams have been proceeding on a roadmap toward a sustainable data archive,\footnote{Funded by a dedicated grant from the Sloan Foundation} designed to protect the legacy of SDSS Data.

Some of the steps on this roadmap, which are currently receiving attention, include:

\begin{itemize}
\item {\bf Archival Quality Storage:}  The Science Archive Server (SAS) file system was not designed to last beyond warranty of the disks, and disk corruption issues require meticulous and time intensive repair. The SAS Team is currently implementing a ceph-based archival quality object storage system \citep{Weil:2006:CSH:1298455.1298485} similar to that used by organizations specializing in big data (e.g. Google and Amazon) providing complete internal redundancy, support for geographical distribution, internal failure detection and self recovery, and inexpensive backup in cloud-based big data object storage systems.
\item {\bf Science Archive Database:}  The census of what is contained on the SAS is managed through a Python system with a database which records the hundreds of millions of file paths, file sizes, and file verification checksums. This system is currently being re-implemented to allow a more seamless and high speed data access.
\item {\bf Migrating the SDSS Software Repository to GitHub:}  The SDSS subversion software repository, currently served along-side the SAS, will be replaced by repositories copied into a GitHub organization (https://github.com/sdss), with GitHub Teams created to manage repository access control, with public release of software including open source licensing, starting with DR15 (e.g. Marvin and the underlying code ``Marvin's Brain").
\item {\bf SDSS Software Framework Development:} The Data and Operations Teams are currently designing a new software framework to provide Python-based tools, including improved data access, database access, data model documentation and machine-readability.
\item {\bf SDSS Software Containers:} Portable images of SDSS systems have been developed and implemented on docker hub, and are currently used at NERSC for the Science Archive Mirror and JHU for Science Archive Webapp development (e.g. Marvin at JHU). The data team is now looking at developing a new wave of such virtual machines to replicate the experience of working on an SDSS computer at the University of Utah.
\end{itemize}

\subsubsection{Modernizing SkyServer}
The SkyServer has been the primary online web portal to the Catalog Archive Server (CAS) since the beginning of SDSS, and although it underwent a significant facelift in 2007, it is now woefully outdated in terms of its layout and the user experience, and generally in terms of its usability and accessibility. The SkyServer has been due for a rewrite with modern web technology for several years now, and we are finally undertaking this daunting task as we wind down SDSS-IV and look forward to SDSS-V. One of the biggest constraints that makes this a difficult enterprise is the large user base that the SkyServer has built over the past 15+ years. We do not want to completely rearrange the site in such a way that users do not recognize it any more, and more importantly, we do not want to break all the functionality that works very well currently in spite of the outdated interface.  In short, we want to adopt a philosophy of going from ``working to working" versions as we modernize the site. We list below the specific changes we are currently working on.
    \begin{itemize}
    \item {\bf Upgrade Technology:} 
First and foremost, we are upgrading the web technology underneath the SkyServer.  This includes everything from the version of HTML and CSS that it was originally written in, to the way that the SkyServer website code is logically organized. We are going toward the MVC (model-view-controller) paradigm that modern websites use to produce modular, reusable and robust web applications.
    \item {\bf Portability:} 
The SkyServer has been a Microsoft Windows application developed with the .NET framework all along.  This has made it very difficult to port even the front end to any other platform.  Now, with the availability of .NET Core, we have the opportunity to migrate the website to a portable platform that can not only run in Linux, but can also be Dockerized.
    \item {\bf Usability:}
 Usability standards have changed considerably since the time when the SkyServer originally came online. In spite of significant upgrades to different parts of the SkyServer over the years, there has not been a comprehensive reexamination of the usability aspects. We aim to rectify this as we redesign the user interface. Usability changes will include effective presentation of information, compatibility with all browsers, responsiveness of page loads, and consistency of display modes (e.g, opening a new tab for results from a query and/or bringing the results pane to the front).
    \item {\bf Accessibility:} 
Accessibility pertains to the versatility of the website and how responsive and easy it is to use and work with for users that are restricted in various ways. This ranges from users on mobile devices to users with restricted access to the internet as well as users with impaired vision or other handicaps. Incorporating modern web design standards and technologies will mostly take care of these aspects, but we will pay special attention to make sure that the SkyServer can be used by as many people as possible anywhere in the world there is internet access.
\item {\bf Integrate SkyServer and Voyages:} 
The SkyServer has an extensive educational sections that contains several levels of classroom exercise based on SDSS data. These are known collectively as the SkyServer Projects.  Voyages is a SkyServer ``spinoff" website that has become quite popular and presents several virtual ``voyages" through the SDSS data for non-scientist audiences. The Voyages website is a much more modern web application that is based on a content management system (CMS) - WordPress. This allows new pages and functionality to be added to Voyages much more easily than SkyServer. As part of the SkyServer modernization, we are migrating all the SkyServer student projects to Voyages and using the same CMS (WordPress) for SkyServer too. We are also integrating Voyages further with SkyServer so that it uses the SkyServer API to run queries on the SDSS data.
\item {\bf Streamline CAS $\Leftrightarrow$ SAS Interface:} 
There are hooks currently between SkyServer/Voyages and the Science Archive Server, but they are awkward at best. The SAS API has recently been upgraded, and the points of access to SAS data that currently exist in SkyServer and Voyages will be updated to use the proper SAS API calls.
\end{itemize}

These steps will help ensure the availability of SDSS data to astronomers for years to come, and long beyond the current funded plans for SDSS-IV and SDSS-V. The CAS data and access tools will at that point be well-positioned to be readily integrated into existing data centers for a minimal incremental cost.

\section{Acknowledgements}

Funding for the Sloan Digital Sky Survey IV has been provided by
the Alfred P. Sloan Foundation, the U.S. Department of Energy Office of
Science, and the Participating Institutions. SDSS-IV acknowledges
support and resources from the Center for High-Performance Computing at
the University of Utah. The SDSS web site is www.sdss.org.

SDSS-IV is managed by the Astrophysical Research Consortium for the 
Participating Institutions of the SDSS Collaboration including the 
Brazilian Participation Group, the Carnegie Institution for Science, 
Carnegie Mellon University, the Chilean Participation Group, the French Participation Group, Harvard-Smithsonian Center for Astrophysics, 
Instituto de Astrof\'isica de Canarias, The Johns Hopkins University, 
Kavli Institute for the Physics and Mathematics of the Universe (IPMU) / 
University of Tokyo, Korean Participation Group, Lawrence Berkeley National Laboratory, 
Leibniz Institut f\"ur Astrophysik Potsdam (AIP),  
Max-Planck-Institut f\"ur Astronomie (MPIA Heidelberg), 
Max-Planck-Institut f\"ur Astrophysik (MPA Garching), 
Max-Planck-Institut f\"ur Extraterrestrische Physik (MPE), 
National Astronomical Observatories of China, New Mexico State University, 
New York University, University of Notre Dame, 
Observat\'ario Nacional / MCTI, The Ohio State University, 
Pennsylvania State University, Shanghai Astronomical Observatory, 
United Kingdom Participation Group,
Universidad Nacional Aut\'onoma de M\'exico, University of Arizona, 
University of Colorado Boulder, University of Oxford, University of Portsmouth, 
University of Utah, University of Virginia, University of Washington, University of Wisconsin, 
Vanderbilt University, and Yale University.

This publication uses data generated via the Zooniverse.org platform, development of which is funded by generous support, including a Global Impact Award from Google, and by a grant from the Alfred P. Sloan Foundation.

This research was made possible through the use of the AAVSO Photometric All-Sky Survey (APASS), funded by the Robert Martin Ayers Sciences Fund.

The Pan-STARRS1 Surveys (PS1) and the PS1 public science archive have been made possible through contributions by the Institute for Astronomy, the University of Hawaii, the Pan-STARRS Project Office, the Max-Planck Society and its participating institutes, the Max Planck Institute for Astronomy, Heidelberg and the Max Planck Institute for Extraterrestrial Physics, Garching, The Johns Hopkins University, Durham University, the University of Edinburgh, the Queen's University Belfast, the Harvard-Smithsonian Center for Astrophysics, the Las Cumbres Observatory Global Telescope Network Incorporated, the National Central University of Taiwan, the Space Telescope Science Institute, the National Aeronautics and Space Administration under Grant No. NNX08AR22G issued through the Planetary Science Division of the NASA Science Mission Directorate, the National Science Foundation Grant No. AST-1238877, the University of Maryland, Eotvos Lorand University (ELTE), the Los Alamos National Laboratory, and the Gordon and Betty Moore Foundation.

This work presents results from the European Space Agency (ESA) space mission Gaia. Gaia data are being processed by the Gaia Data Processing and Analysis Consortium (DPAC). Funding for the DPAC is provided by national institutions, in particular the institutions participating in the Gaia MultiLateral Agreement (MLA). The Gaia mission website is \url{https://www.cosmos.esa.int/gaia}. The Gaia archive website is \url{https://archives.esac.esa.int/gaia}.

We would like to thank the e-Science Institute at the University of Washington, Seattle for their hospitality during DocuVana 2018. This event held in May 2018 kick started the documentation for DR15 (including this paper) and was organized by Jennifer Sobeck and Jos\'e S\'anchez-Gallego and Anne-Marie Weijmans and attended by Amy Jones, Ben Murphy, Bonnie Souter, Brian Cherinka, David Stark, David Law, Dan Lazarz, Gail Zasowski, Joel Brownstein, Jordan Raddick, Julie Imig, Karen Masters, Kyle Westfall, Maria Argudo-Fern\'andez, Michael Talbot, Rachael Beaton, Renbin Yan and Sten Hasselquist (as well as Becky Smethurst, Rita Tojeiro, Ben Weaver and Ani Thaker via video link). 

This research made use of \textsc{astropy}, a community-developed core \textsc{python} ({\tt http://www.python.org}) package for Astronomy \citep{2013A&A...558A..33A}; \textsc{ipython} \citep{PER-GRA:2007}; \textsc{matplotlib} \citep{Hunter:2007}; \textsc{numpy} \citep{:/content/aip/journal/cise/13/2/10.1109/MCSE.2011.37}; \textsc{scipy} \citep{citescipy}; and \textsc{topcat} \citep{2005ASPC..347...29T}.

\bibliography{references}{}
\bibliographystyle{aasjournal}

\end{document}